\documentclass[a4paper,fleqn]{cas-dc}

\usepackage[authoryear]{natbib}

\def\tsc#1{\csdef{#1}{\textsc{\lowercase{#1}}\xspace}}
\tsc{WGM}
\tsc{QE}
\tsc{EP}
\tsc{PMS}
\tsc{BEC}
\tsc{DE}

\usepackage[authoryear]{natbib} 
\usepackage{algorithm}
\usepackage{changepage}
\usepackage{algorithmic}
\usepackage{subcaption}
\usepackage[most]{tcolorbox}
\usepackage{booktabs} 
\usepackage{tabularx} 
\usepackage{array}

\usepackage{array} 
\newcolumntype{Y}{>{\raggedright\arraybackslash\hspace{0pt}}X} \newcolumntype{L}[1]{>{\raggedright\arraybackslash\hspace{0pt}}p{#1}} \newcolumntype{C}[1]{>{\centering\arraybackslash}p{#1}}

\usepackage{listings}
\usepackage{url}
 
\author{
    \textbf{Rui Jia}\textsuperscript{1}, 
    \textbf{Min Zhang}\textsuperscript{1,$\ddagger$,$\dagger$}, 
    \textbf{Fengrui Liu}\textsuperscript{1},
    \textbf{Bo Jiang}\textsuperscript{1}, 
    \textbf{Kun Kuang}\textsuperscript{2}, \textbf{Zhongxiang Dai}\textsuperscript{3} \\
    \textsuperscript{1}East China Normal University,
    \textsuperscript{2}Zhejiang University, \\
    \textsuperscript{3}The Chinese University of Hong Kong, Shenzhen
    \\
    \small \textsuperscript{$\ddagger$}Project leader \quad 
    \href{52285901028@stu.ecnu.edu.cn}{52285901028@stu.ecnu.edu.cn} \quad
    \textsuperscript{$\dagger$}Corresponding author: \href{mailto:mzhang@cs.ecnu.edu.cn}{mzhang@cs.ecnu.edu.cn}
}

\begin{document}
\let\WriteBookmarks\relax
\def\floatpagepagefraction{1}
\def\textpagefraction{.001}
\shorttitle{EduAgentQG for Personalized Question Generation}

\title[mode=title]{EduAgentQG: Multi-Agent Personalized Mathematics Question Generation with Explicit Diversity and Objective-Aware Evaluation}
 
\begin{abstract}
In intelligent education, personalized mathematics question generation aims to produce mathematics questions that satisfy educational requirements while supporting adaptive assessment and learning. Existing LLM-based single-agent and multi-agent methods improve generation flexibility, but they still tend to rely on aggregated feedback or model randomness, making it difficult to jointly ensure dimension-wise objective alignment and controllable diversity. To address these challenges, we propose \textbf{EduAgentQG}, a multi-agent collaborative framework for personalized mathematics question generation with explicit diversity and objective-aware evaluation. EduAgentQG organizes question generation as a closed-loop process of planning, writing, evaluation, refinement, and checking: structured generation plans and multiple generation directions guide candidate generation, while fine-grained evaluation verifies logical correctness, solvability, and objective alignment in knowledge concepts, difficulty, grade level, and core competencies. We first construct a mathematics question generation benchmark containing 10,273 questions across Grades 1-9, covering 634 knowledge concepts, 16 core competencies, and three difficulty levels; for evaluation, it is organized into two subsets: MathChoice, with 489 educational objectives for multiple-choice question generation, and MathBlank, with 500 educational objectives for fill-in-the-blank question generation. Experiments using Gemini-2.5-Flash, GPT-4o-mini, and Qwen2.5-72B-Instruct show that EduAgentQG consistently outperforms COT, COT$_N$, ReAct, and EQPR in diversity, Objective Consistency, and Win Rate. On Gemini-2.5-Flash, EduAgentQG improves Win Rate by 3.90 and 4.75 percentage points on MathChoice and MathBlank, over the strongest baseline, demonstrating its ability to generate diverse, solvable, and educationally aligned personalized mathematics questions with stronger controllability. The code is available at \url{https://github.com/ECNU-RAIL/EduAgentQG-IPM2026}.
\end{abstract}


\maketitle

\section{Introduction}

\begin{figure*}[tb]
    \centering
    \includegraphics[width=\linewidth]{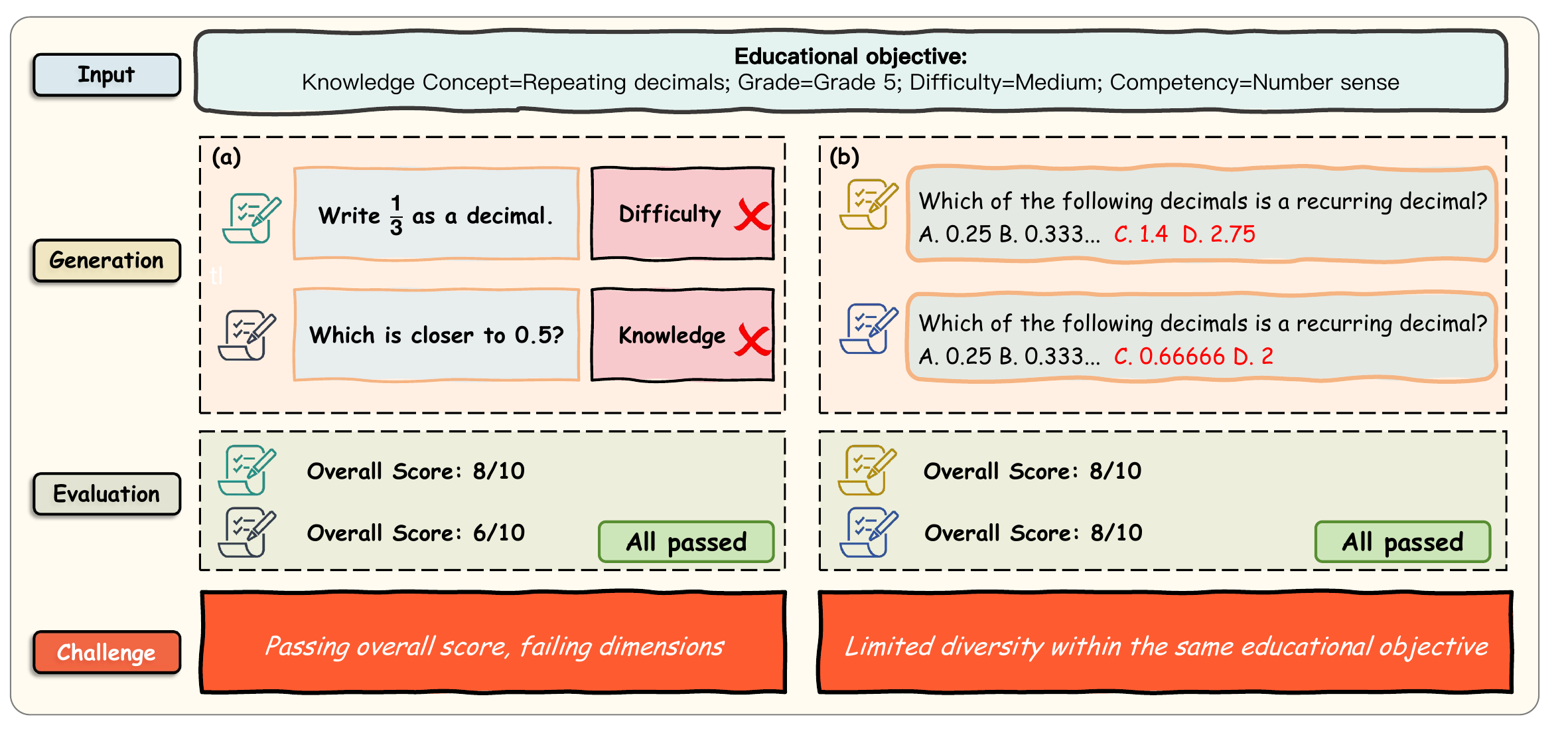}
    \caption{Two major challenges in existing question generation methods. (a) Overall scoring may obscure deficiencies in specific educational dimensions, allowing questions that fail knowledge, difficulty, or competency requirements to pass evaluation. (b) Questions generated under the same educational objective often exhibit high similarity in semantic content and assessment perspective, resulting in limited diversity.}
    \label{fig:intro}
\end{figure*}

Personalized Question Generation (QG) aims to automatically generate questions that align with user-specified educational objectives, which specify knowledge coverage, difficulty requirements, and core competency development~\citep{hang2024mcqgen,hwang2024contextualized,wang2021math,jiang2024improving}. In practical teaching, educators typically engage in iterative cycles of planning, drafting, validation, and optimization to design questions that support both assessment and cognitive development~\citep{rembert2020examining,walkington2017designing,surya2020development}. During this process, teachers need to balance conceptual understanding, core competencies, and difficulty levels to ensure that questions are both diagnostically effective and pedagogically meaningful~\citep{kliebard1970, wiggins2005}. However, manually creating high-quality questions for diverse learning needs is time-consuming and difficult to scale. Although teachers can identify specific learning needs and define educational objectives, generating questions that precisely match these objectives within limited time remains challenging~\citep{rembert2019exploring,ozgen2021checklist,ikashaum2023exion}. Therefore, automated mathematics question generation based on predefined educational objectives has emerged as an important approach to reducing teachers' cognitive load and time costs, improving instructional efficiency, and enhancing the scalability of educational resources.

Traditional question generation methods mainly include template-based, neural generation, and knowledge-enhanced approaches. Template-based methods rely on handcrafted rules and grammar patterns, offering controllable structures but limited flexibility and diversity~\citep{nandhini2011, lu2021}. Neural sequence-to-sequence methods improve fluency by learning generation patterns from data~\citep{xu2021procedural,dave2021math}, but they still struggle to capture mathematical structures and reasoning processes~\citep{zhou2023}. Knowledge-enhanced and structure-aware methods further incorporate mathematical structures or external knowledge to improve logical plausibility and structural coherence~\citep{qin2023}. More recently, large language models have brought substantial improvements to question generation~\citep{chudziak2025ai,beauchamp2025using,liu2025one}.

Many LLM-based QG methods can be viewed as following a three-stage process: in the \textbf{input stage}, educational objectives are provided; in the \textbf{generation stage}, models create questions based on these objectives; and in the \textbf{evaluation stage}, generated questions are assessed, with only satisfactory candidates retained as final outputs. Within this framework, existing approaches can be broadly categorized into single-agent and multi-agent methods. Single-agent methods rely on one model to perform generation, reasoning, and self-evaluation within a unified process~\citep{pham2024chatgpt,li2024generating,li2024large,liang2025chain,shah2024ai,lee2024math}. These methods are simple and flexible, but the same model must simultaneously handle planning, generation, and quality control, which limits fine-grained control over educational constraints and diversity. In contrast, multi-agent methods decompose the generation process into multiple roles, such as generation, evaluation, critique, and refinement, allowing different agents to collaborate for more reliable outputs~\citep{wang2025llm,yan2025mathagent,gao2025agent4edu,cheng2025,karbasi2025,manem2025sand}. However, existing multi-agent methods still tend to rely on aggregated feedback or implicit reasoning patterns, and often lack explicit mechanisms for dimension-wise educational evaluation and controllable diversity.

Despite these advances, existing methods still face two major challenges, as illustrated in Figure~\ref{fig:intro}. First, in the \textbf{evaluation stage}, current scoring mechanisms generally rely on overall scores, meaning that questions may be accepted even when they fail to satisfy certain educational dimensions, such as knowledge concept coverage, difficulty alignment, or competency requirements (Figure~\ref{fig:intro}a). This limitation arises because overall scores tend to obscure deficiencies in specific educational aspects, allowing questions that appear acceptable at a coarse level to pass evaluation. Second, in the \textbf{generation stage}, question diversity is often limited; questions generated for the same \textbf{educational objective} tend to be highly similar in semantic content and assessment perspective (Figure~\ref{fig:intro}b). This is because existing methods largely rely on the inherent randomness of large models and lack explicit mechanisms to control diversity.

To address these challenges, we propose \textbf{EduAgentQG}, a multi-agent collaborative framework for personalized mathematics question generation. The key idea of EduAgentQG is to jointly support direction-guided diversity control and multi-dimensional objective-aware evaluation. Rather than relying mainly on model randomness to obtain varied outputs, the \textbf{Planner} transforms each educational objective into structured design plans with multiple generation directions, and the \textbf{Writer} generates and iteratively revises candidate questions with different contexts, presentation styles, and reasoning paths. Rather than relying on an aggregated quality score, the \textbf{Solver} and \textbf{Educator} conduct complementary fine-grained validation across logical and educational dimensions: the \textbf{Solver} examines logical soundness, solvability, and ambiguity, while the \textbf{Educator} checks grade-level, knowledge-concept, difficulty, and core-competency alignment. The \textbf{Checker} then performs final verification of correctness and clarity. Through this design, EduAgentQG goes beyond ordinary multi-agent role division by explicitly linking generation diversity and objective verification to the structured educational objective, enabling controllable diversity and dimension-wise quality control for personalized mathematics question generation.

\subsection{Research Objectives}

Motivated by the above challenges, this study aims to develop a controllable framework for personalized mathematics question generation that can generate questions with both educational alignment and diversity. Specifically, this study focuses on the following research objectives:

\begin{itemize}
    \item To investigate how educational objectives can be transformed into structured generation guidance that supports alignment with knowledge concepts, difficulty levels, grade requirements, and core competencies.

    \item To explore how controllable diversity and dimension-wise quality control can be jointly achieved under the same educational objective, so that generated questions are both pedagogically aligned and meaningfully varied.

    \item To evaluate whether the proposed multi-agent framework can consistently improve mathematics question quality, diversity, and Objective Consistency across different question formats and backbone models.
\end{itemize}

\subsection{Contributions}
The main contributions are summarized as follows:

\begin{itemize}

\item We propose EduAgentQG, an objective-aware multi-agent framework for personalized mathematics question generation. The framework coordinates planning, question generation, evaluation, and refinement to generate questions that satisfy structured educational objectives while improving both question diversity and Objective Consistency.

\item We design a planning-based diversity control mechanism. Given the same educational objective, the Planner derives multiple generation directions that guide the Writer to generate questions with diverse contextual settings, presentation styles, and reasoning paths while maintaining alignment with the target educational objective.

\item We develop a fine-grained objective-aware evaluation mechanism based on complementary Solver and Educator agents. The mechanism jointly verifies logical validity and pedagogical alignment by evaluating solvability, ambiguity, grade-level alignment, knowledge concept alignment, difficulty level, and core competency alignment, providing multi-dimensional feedback for iterative question refinement.

\item Extensive experiments on the MathChoice and MathBlank benchmarks demonstrate that EduAgentQG consistently outperforms representative baseline methods in diversity, Objective Consistency, and Win Rate across different backbone models and question formats.

\end{itemize}

\section{Related Works}

\subsection{Question Generation and Diversity}

Question generation has evolved from template-based and neural generation methods to LLM-based single-agent generation and, more recently, multi-agent collaboration. Early approaches relied on handcrafted templates and grammar rules, which provided controllable structures and logical consistency but limited flexibility, scalability, and diversity~\citep{nandhini2011, lu2021}. With the development of deep learning, neural sequence-to-sequence models became widely used by learning generation patterns from data~\citep{xu2021procedural,dave2021math}. These methods improved surface fluency and reduced the dependence on manually designed templates, yet they still struggled to capture mathematical structures, multi-step reasoning, and constraint satisfaction in educational scenarios~\citep{zhou2023}. To improve the logical and structural quality of generated questions, knowledge-enhanced and structure-aware methods were further explored. For example, \citet{qin2023} proposed MaPKG, which incorporates expression trees and commonsense knowledge graphs to enhance logical plausibility and structural coherence. Nevertheless, these methods still largely depend on predefined structures, external resources, or learned generation patterns, making it difficult to explicitly control question diversity under the same educational objective.

Recent studies have further explored integrated automatic question generation systems that combine question generation with educational content analysis. \citet{saleem2026information} proposed Questify-TheEduBot, a unified NLP-based framework that integrates transformer-based models, topic modeling, sentiment analysis, keyword extraction, and semantic similarity to generate multiple types of educational questions, including multiple-choice, cloze, and descriptive questions. This line of work shows the value of combining question generation with auxiliary analysis modules to improve pedagogical relevance and interpretability. However, such systems mainly focus on generating questions from educational content and validating general content-level properties, while the problem of generating diverse questions under explicit educational objectives remains less explored.

The emergence of large language models has enabled more flexible single-agent question generation frameworks. \citet{li2024large} proposed a Chain-of-Thought-based single-agent approach, allowing step-by-step reasoning during generation. Compared with template-based methods, such LLM-based generation improves flexibility and can produce more diverse surface forms. However, because a single model is usually responsible for planning, generation, reasoning, and self-evaluation simultaneously, it is difficult to ensure stable control over educational constraints, logical correctness, and diversity. In particular, without systematic evaluation and feedback-driven optimization, generated questions may still suffer from insufficient difficulty alignment, incomplete reasoning, or limited pedagogical diversity.

To further improve generation quality and reliability, multi-agent collaboration has been explored. \citet{karbasi2025} proposed the Teacher-Critic-CEO framework, where three agents collaboratively generate and refine questions to enhance overall quality. \citet{cheng2025} introduced EQPR, which combines a generation agent and an evaluation agent with Monte Carlo Tree Search to dynamically control the generation process and produce more stable outputs. Beyond educational question generation, role-specialized multi-agent frameworks have also been investigated for complex question answering tasks. \citet{wang2026datafactory} proposed DataFactory, a collaborative multi-agent framework for advanced table question answering, where a Data Leader coordinates Database and Knowledge Graph teams to decompose complex queries into structured and relational reasoning subtasks. These studies show that assigning different roles to different agents can improve reliability by introducing planning, coordination, evaluation, and refinement into complex reasoning pipelines. However, existing multi-agent methods still do not explicitly decompose educational objectives into multiple generation directions, and their diversity often depends on model randomness or implicit reasoning patterns rather than controllable planning.

Overall, many existing generation methods, especially LLM-based ones, still rely on model randomness or implicit reasoning patterns to achieve diversity, often yielding semantically or structurally similar questions under the same educational objective. In contrast, our work enhances diversity by specifying multiple generation directions in the planning stage and performing multi-perspective rewriting in the subsequent optimization stage, thereby systematically increasing meaningful variation throughout the entire generation process.

\subsection{Question Evaluation and Optimization}

Evaluation and optimization are critical for ensuring logical soundness, educational alignment, and overall question quality. Early methods mainly relied on automated text similarity metrics such as BLEU and ROUGE~\citep{lu2021}. Although these metrics are easy to compute and useful for measuring lexical overlap, they do not adequately capture reasoning completeness, stepwise logic, solvability, or pedagogical relevance. This limitation is particularly important in mathematics question generation, where a generated question may appear fluent but still fail to satisfy the intended knowledge concept, difficulty level, or reasoning requirement. More broadly, \citet{martinez2026assessing} also noted that general metrics such as BLEU, METEOR, ROUGE, and BERTScore mainly capture lexical or semantic overlap with reference texts, and may fail to reflect the appropriateness of valid but diverse outputs in open-ended generation settings. This further suggests that educational question generation requires task-specific and dimension-aware evaluation beyond surface-level similarity.

Later works introduced more task-oriented evaluation strategies. Some studies used solver-based verification to check reasoning chains and answer correctness~\citep{zhou2023}, while others compared generated questions with curriculum standards or benchmark question banks to assess educational relevance~\citep{karbasi2025}. These methods improve evaluation reliability by considering correctness and curriculum alignment beyond surface-level text similarity. However, they often provide coarse-grained or aggregated feedback, which makes it difficult to identify which specific educational dimension causes a generated question to fail.

Recent work on knowledge-enhanced LLM reasoning also highlights the importance of structured external knowledge for improving reliability and interpretability. \citet{gong2026enhancing} proposed MGPrompt for knowledge graph question answering, which integrates entity-level, relation-level, and subgraph-level knowledge with structured reasoning path-augmented prompting. Although knowledge graph question answering differs from educational question generation, this line of research supports the broader motivation for using structured knowledge and retrieval mechanisms to guide LLM-based reasoning and reduce unreliable outputs.

Recently, multi-agent collaborative frameworks have adopted a generate--evaluate--optimize cycle. The EQPR framework~\citep{cheng2025} employs a generation agent and an evaluation agent, with Monte Carlo Tree Search (MCTS) dynamically controlling whether to output or regenerate questions based on feedback, thereby enhancing both question quality and educational alignment. Similarly, \citet{wang2024} proposed a multi-agent multiple-choice generation system that relies on evaluation agents to make multi-dimensional decisions without an explicit planning stage, achieving notable quality improvements. These approaches demonstrate the value of iterative evaluation and refinement, but their feedback mechanisms are still often aggregated or implicitly represented, limiting strict control over individual educational constraints.

However, many existing methods still primarily depend on aggregated or implicit feedback, which limits strict control over knowledge coverage, difficulty alignment, competency alignment, and solvability. Our multi-agent evaluation framework uses fine-grained binary judgments for each dimension and accepts questions only when all strict criteria are satisfied. This design provides stronger quality control and interpretability while ensuring pedagogical and logical reliability.

\begin{figure*}[tb]
    \centering
    \includegraphics[width=\linewidth]{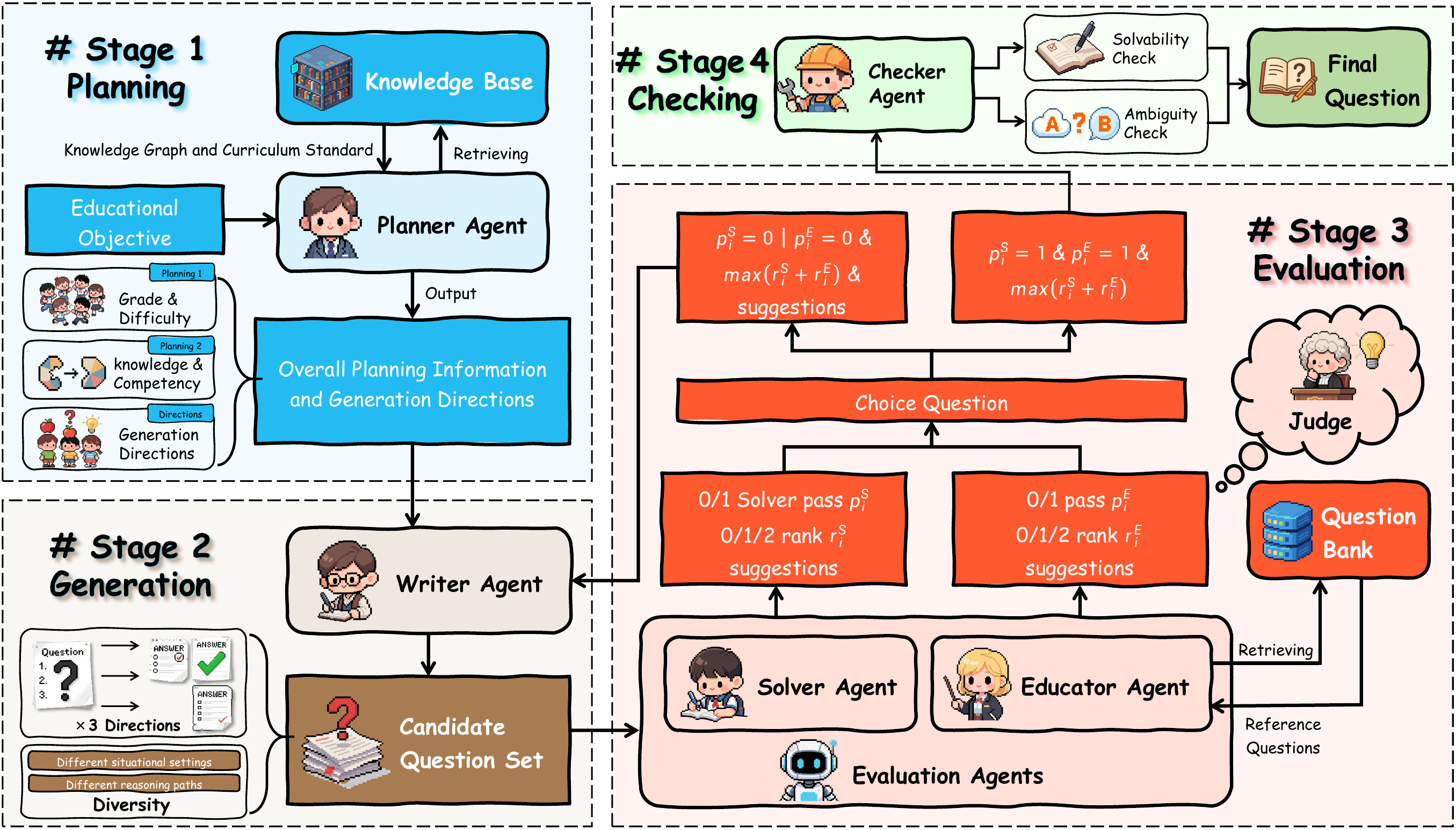}
    \caption{Overview of the EduAgentQG multi-agent framework. The Planner defines generation directions, the Writer produces diverse candidate questions, the Solver and Educator evaluate each question along multiple dimensions, and the Checker ensures final correctness and clarity.}
    \label{fig:framework}
\end{figure*}

\section{Methodology}

\subsection{Task Definition}

Automatically generating mathematics questions that conform to a specific educational objective is challenging because multiple constraints must be satisfied simultaneously. A valid question should align with the target grade level, cover the required knowledge concepts, match the intended difficulty level, and reflect the specified core competency. In addition, the question should be logically complete, unambiguous, and solvable, with a unique correct answer. To support educational diversity, repeated generation under the same educational objective should produce questions that differ in context, phrasing, or reasoning paths.

Formally, an educational objective is represented as
\begin{equation}
E = (G, K, D, S),
\end{equation}
where $G$ denotes the target grade level ($1 \leq G \leq 9$), $K$ denotes the set of knowledge concepts, $D \in \{\text{easy}, \text{medium}, \text{hard}\}$ specifies the target difficulty level, and $S$ represents the target core competency dimension. Specifically, $G$ is determined by the school grade in which the corresponding mathematical content is taught. $K$ refers to the mathematical knowledge concepts required to solve the question, such as reading and writing numbers within 100 million, fractions, quadrilaterals, compound statistical tables, or 24-hour time notation; a question may involve one or more knowledge concepts. $D$ reflects the expected cognitive and reasoning demand of the question. Easy questions usually require direct application of a single concept or routine procedure with simple computation. Medium questions usually involve multi-step reasoning, moderate transformation of given conditions, or the integration of related concepts. Hard questions usually require complex reasoning, non-routine problem solving, implicit condition identification, or the coordinated use of multiple concepts. $S$ denotes the target core competency dimension reflected by the question, such as Computation, geometric intuition, Reasoning, number sense, or application awareness. Together, these four components define the educational objective that the generated question should satisfy.

Given an educational objective $E$, the task of EduAgentQG is to generate a single question
\begin{equation}
q = (\text{stem}, \text{answer}),
\end{equation}
where \textit{stem} denotes the question text, which may include options for multiple-choice questions, and \textit{answer} denotes the unique correct solution. The generated question $q$ must satisfy the constraints imposed by $E$, ensure logical soundness and answer correctness, and allow controlled diversity across repeated generations under the same educational objective.

\begin{algorithm}[tb]
\small
\caption{EduAgentQG Question Generation Procedure}
\label{alg:eduagentqg}
\begin{algorithmic}[1]
\REQUIRE educational objective $E=(G,K,D,S)$, knowledge base $\mathcal{B}$, 
question bank $\mathcal{Q}_{\mathrm{bank}}$, maximum rewriting steps 
$T_{\mathrm{rewrite}}$, maximum restart attempts $T_{\mathrm{restart}}$
\ENSURE final generated question $q^{*}$ if at least one candidate passes all 
required checks; otherwise \textsc{Failure}

\STATE Initialize $\mathcal{Q}_{\mathrm{pass}} \gets \emptyset$

\FOR{$a = 1$ to $T_{\mathrm{restart}}$}
    \STATE \textbf{Planning:} $(Plan, Dir) \gets Planner(E,\mathcal{B})$
    \STATE \textbf{Generation:} $\mathcal{Q}^{(0)} \gets Writer(Plan, Dir)$

    \FOR{$t = 0$ to $T_{\mathrm{rewrite}}$}

        \IF{$t < T_{\mathrm{rewrite}}$}
            \STATE Initialize $\mathcal{Q}^{(t+1)} \gets \emptyset$
        \ENDIF

        \FOR{each $q_i^{(t)} \in \mathcal{Q}^{(t)}$}
            \STATE \textbf{Evaluation:} $(p_i^\mathrm{S}, r_i^\mathrm{S}, 
            \mathcal{F}_i^\mathrm{S}) \gets Solver(q_i^{(t)})$
            \STATE $(p_i^\mathrm{E}, r_i^\mathrm{E}, \mathcal{F}_i^\mathrm{E}) 
            \gets Educator(q_i^{(t)}, E, \mathcal{Q}_{\mathrm{bank}})$

            \IF{$p_i^\mathrm{S} = 1$ \textbf{and} $p_i^\mathrm{E} = 1$}
                \STATE $\mathcal{Q}_{\mathrm{pass}} \gets 
                \mathcal{Q}_{\mathrm{pass}} \cup \{q_i^{(t)}\}$
            \ELSE
                \IF{$t < T_{\mathrm{rewrite}}$}
                    \STATE \textbf{Refinement:} $q_i^{(t+1)} \gets 
                    Writer(q_i^{(t)}, \mathcal{F}_i^\mathrm{S}, 
                    \mathcal{F}_i^\mathrm{E})$
                    \STATE $\mathcal{Q}^{(t+1)} \gets 
                    \mathcal{Q}^{(t+1)} \cup \{q_i^{(t+1)}\}$
                \ENDIF
            \ENDIF
        \ENDFOR

        \IF{$\mathcal{Q}_{\mathrm{pass}} \neq \emptyset$}
            \STATE \textbf{break}
        \ENDIF

        \IF{$t = T_{\mathrm{rewrite}}$}
            \STATE \textbf{break}
        \ENDIF

        \IF{$\mathcal{Q}^{(t+1)} = \emptyset$}
            \STATE \textbf{break}
        \ENDIF
    \ENDFOR

    \IF{$\mathcal{Q}_{\mathrm{pass}} \neq \emptyset$}
        \STATE \textbf{break}
    \ENDIF
\ENDFOR

\IF{$\mathcal{Q}_{\mathrm{pass}} = \emptyset$}
    \STATE \RETURN \textsc{Failure}
\ENDIF

\STATE \textbf{Selection:} $\hat{q} \gets 
\arg\max_{q_i \in \mathcal{Q}_{\mathrm{pass}}}
(r_i^\mathrm{S} + r_i^\mathrm{E})$
\STATE \textbf{Checking:} $q^{*} \gets Checker(\hat{q})$
\STATE \RETURN $q^{*}$

\end{algorithmic}
\end{algorithm}

\subsection{Multi-Agent Personalized Question Generation with Explicit Diversity and Objective-Aware Evaluation}

We propose EduAgentQG, an objective-aware multi-agent framework for personalized mathematics question generation that aims to jointly improve controllable diversity and Objective Consistency under explicit educational objectives. The framework consists of five specialized agents, namely the Planner, Writer, Solver, Educator, and Checker, which are arranged in a structured and sequential generation pipeline, as illustrated in Figure~\ref{fig:framework}. Different from standard generate-and-evaluate pipelines, EduAgentQG is organized around two core mechanisms: generation-direction-based diversity control and fine-grained objective-aware evaluation. Specifically, the Planner transforms the educational objective into structured planning information and generation directions, the Writer generates and revises candidate questions under these directions, the Solver evaluates logical clarity and solvability, the Educator examines objective alignment, and the Checker performs final verification before output.

EduAgentQG addresses these challenges through two linked mechanisms. The Planner decomposes the educational objective into structured planning information and multiple generation directions, which guide the Writer to produce candidate questions with varied contexts, presentation styles, and reasoning structures. The Solver and Educator then perform complementary dimension-wise validation: the Solver checks logical clarity, solvability, and ambiguity, whereas the Educator evaluates grade-level, knowledge-concept, difficulty, and core-competency alignment. A candidate is retained only when it passes both agents; otherwise, the Writer revises it using their feedback. Thus, diversity is explicitly planned rather than obtained through random sampling, and objective failures cannot be masked by an aggregated score.

\textbf{Planner Agent.}
The generation process begins with the Planner, which analyzes the input educational objective $E$ and retrieves relevant curriculum standards and knowledge graph information through a RAG mechanism. The retrieved curriculum standards provide the grade-level scope, learning requirements, and competency expectations used for planning. Since these standards are provided as external retrieval resources rather than fixed model parameters, they can be replaced or updated according to different countries, regions, textbook systems, or school curricula. Therefore, EduAgentQG does not assume a universal grade-level curriculum; instead, grade-level alignment is judged with respect to the curriculum standards supplied to the RAG module. Based on these resources, the Planner decomposes $E$ into a clear and interpretable structured design plan:
\begin{equation}
P(E, \text{RAG}) = (Plan, Dir).
\end{equation}
The component \textit{Plan} contains three essential planning elements:
(1) a \textit{knowledge concept plan} that identifies the concepts to be addressed, prerequisite relations, and common misconceptions;
(2) a \textit{difficulty plan} tailored to the cognitive characteristics of the target grade level;
and (3) a \textit{competency plan} specifying how the target core competency should be reflected.
The component \textit{Dir} consists of high-level generation directions, each representing a distinct context, presentation style, or reasoning structure. The Planner does not directly produce concrete questions; instead, it provides structured and constraint-driven guidance for downstream agents.

\textbf{Writer Agent.}  
Using $(Plan, Dir)$, the Writer generates an initial set of $N$ candidate questions:
\begin{equation}
\mathcal{Q}^{(0)} = \{ q_1^{(0)}, q_2^{(0)}, \dots, q_N^{(0)} \}.
\end{equation}
Each candidate question $q_i$ consists of a question stem and its unique correct answer. During generation, the Writer follows the knowledge concept plan, difficulty plan, and competency plan defined in \textit{Plan}. For each direction $dir_i \in Dir$, it adapts the question context, structure, or solution approach, thereby enabling controlled diversity. If a question fails evaluation, the Writer revises it using explicit feedback from the evaluation agents, allowing the system to explore alternative phrasings and solution paths while preserving the underlying educational objective.

\textbf{Solver Agent.}  
The Solver evaluates logical clarity, solvability, and instructional validity. Specifically, it checks whether the question conditions are sufficient, whether the question avoids ambiguity or unintentional hints, and whether the solution path is appropriate for the target grade level. For each question $q_i$, the Solver produces a binary pass signal:
\begin{equation}
p_i^\mathrm{S} =
\begin{cases}
1, & q_i \in \mathcal{C}_{\mathrm{logic}}, \\
0, & q_i \notin \mathcal{C}_{\mathrm{logic}}.
\end{cases}
\end{equation}
The Solver also assigns a ranking score $r_i^\mathrm{S}$ that reflects the logical robustness and pedagogical soundness of the question in typical classroom settings.

\textbf{Educator Agent.}  
The Educator examines whether a question aligns with the target knowledge concepts $K$, difficulty level $D$, grade level $G$, and core competency requirement $S$. Because difficulty and competency alignment can be subtle, the Educator employs a RAG mechanism to retrieve a reference set
$\mathcal{Q}_{\mathrm{ref}} \subset \mathcal{Q}_{\mathrm{bank}}$ of representative easy, medium, and hard questions associated with the same knowledge concept and grade level, as well as comparable core competency requirements. These reference questions serve as anchors for evaluating a candidate question's difficulty and competency alignment. The Educator outputs:
\begin{equation}
p_i^\mathrm{E} =
\begin{cases}
1, & q_i \in \mathcal{C}_{\mathrm{edu}}(G, K, D, S), \\
0, & q_i \notin \mathcal{C}_{\mathrm{edu}}(G, K, D, S).
\end{cases}
\end{equation}
and a ranking score $r_i^\mathrm{E}$. Questions that pass both the Solver and Educator form the set:
\begin{equation}
\mathcal{Q}_{\mathrm{pass}} = \{ q_i \mid p_i^\mathrm{S}=1 \wedge p_i^\mathrm{E}=1 \}.
\end{equation}
\textbf{Checker Agent.}  
The Checker performs final verification by ensuring answer correctness and eliminating ambiguity. After the iterative evaluation and rewriting process, EduAgentQG selects the final candidate from the set of questions that pass both the Solver and Educator. If $\mathcal{Q}_{\mathrm{pass}}$ is non-empty, the candidate with the highest combined Solver--Educator score is selected:
\begin{equation}
\hat{q} =
\arg\max_{q_i \in \mathcal{Q}_{\mathrm{pass}}}
(r_i^\mathrm{S} + r_i^\mathrm{E}).
\end{equation}
If multiple candidates share the same combined score, the one with the higher Educator score is preferred. If no candidate passes both agents after the maximum rewriting steps, EduAgentQG restarts the generation process with new planning information and candidate questions rather than directly outputting an unpassed candidate. This ensures that the final selection is always made from $\mathcal{Q}_{\mathrm{pass}}$. Once $\hat{q}$ is selected, it is passed to the Checker, and the final output is obtained as:
\begin{equation}
q^* = \text{Checker}(\hat{q}).
\end{equation}

\textbf{Iterative Closed-Loop Refinement.}  
Not all questions pass evaluation on the first attempt. For each failing question, feedback from the Solver and Educator, denoted as $\mathcal{F}(q_i)$, including failure reasons and revision suggestions, is used by the Writer to generate a revised version:
\begin{equation}
q_i^{(t+1)} = \text{Writer}\big(q_i^{(t)}, \mathcal{F}(q_i^{(t)})\big).
\end{equation}
This refinement is performed iteratively:
\begin{equation}
t = 0, 1, \dots, T_{\mathrm{rewrite}} - 1.
\end{equation}

Across iterations, candidate questions evolve toward stronger logical soundness, improved educational alignment, and richer diversity in context or reasoning. Candidates that pass both the Solver and Educator are added to $\mathcal{Q}_{\mathrm{pass}}$, whereas candidates that fail either evaluation are revised by the Writer according to the corresponding feedback. If no candidate enters $\mathcal{Q}_{\mathrm{pass}}$ after the maximum number of rewriting steps, the framework restarts the planning and generation process. Once $\mathcal{Q}_{\mathrm{pass}}$ is non-empty, the combined Solver--Educator ranking scores are used to select the final candidate from the passed set. The overall multi-agent workflow of EduAgentQG is summarized in Algorithm~\ref{alg:eduagentqg}.

\section{Experiments}

\subsection{Dataset Construction and Annotation}

We constructed a mathematics question dataset from teacher-provided mathematics questions collected from primary and junior secondary schools. The raw questions covered school-level mathematics resources across Grades 1--9. Before annotation, we organized the collected questions by grade level and question type, and removed duplicated, incomplete, or ill-formed items. We retained only questions with clear stems, sufficient conditions, and unique answers, so that each item could be reliably associated with an explicit educational objective.

After preprocessing, the dataset contains 10,273 mathematics questions in total. Each question is associated with a structured educational objective defined by four dimensions: grade level, knowledge concept, difficulty level, and core competency. The dataset covers 634 knowledge concepts, 16 core competency categories, and three difficulty levels, with questions sampled across different grade levels, mathematical topics, competency categories, and question formats. This design allows the benchmark to evaluate not only whether a model can generate mathematically valid questions, but also whether it can follow fine-grained educational requirements.

The grade-level labels were determined with reference to the \textit{Mathematics Curriculum Standards for Compulsory Education (2022 Edition)} issued by the Ministry of Education of the People's Republic of China. This standard defines the educational goals, content requirements, academic requirements, and assessment guidance for mathematics in compulsory education from Grades 1--9. In this study, each question was assigned to a grade level according to the curriculum stage and the grade-level progression of the corresponding knowledge concept in the source question bank. Therefore, the grade labels should be interpreted within the context of the Chinese compulsory education mathematics curriculum rather than as universal grade-level standards. Since grade-level requirements may vary across countries, regions, and textbook systems, EduAgentQG provides curriculum information to the Planner through the RAG-based knowledge base, allowing the framework to be adapted to other curriculum systems by replacing the underlying curriculum resources.

To ensure annotation reliability, 30 master's and doctoral students majoring in mathematics education participated in the annotation process. Each question was independently annotated by two annotators, and any disagreement was resolved by a third annotator. During annotation, grade level was determined according to the curriculum position of the mathematical content, and knowledge concepts were assigned based on the concepts required to solve the question. Difficulty labels were determined by considering the expected cognitive and reasoning demand, including the number of solution steps, the degree of condition transformation, the level of concept integration, and whether the question required routine application or non-routine problem solving.

Core competency labels were assigned according to the mathematical abilities required to solve the question. The definitions of the 16 core competency categories are provided in Appendix~C. A question may be assigned either a single competency label or multiple competency labels when solving it requires more than one competency dimension. For example, a geometry problem involving both spatial representation and logical justification may be annotated with Spatial concept and Reasoning. When annotators disagreed on competency labels, the disagreement was resolved by a third annotator according to the competency definitions and the dominant mathematical abilities required by the question. If multiple competencies were judged to be essential rather than conflicting, all relevant labels were retained.

The final annotation of each question is represented as $(G, K, D, S)$, where $G$ denotes the grade level, $K$ denotes the knowledge concept specification, $D$ denotes the difficulty level, and $S$ denotes the core competency specification. Both $K$ and $S$ may contain one or more labels. Compared with datasets that provide only topic-level labels, this annotation protocol links each question to an explicit and fine-grained educational objective, making it possible to evaluate objective-aware question generation in a more controlled manner. To provide a concrete illustration of the annotation protocol, Appendix~\ref{app:annotation_examples} presents five representative examples with their question texts, answers, grade levels, knowledge concepts, difficulty levels, and competency labels.

For evaluation, we divide the benchmark into two format-specific subsets according to question type: \textit{MathChoice} and \textit{MathBlank}. \textit{MathChoice} contains 489 multiple-choice evaluation instances, and \textit{MathBlank} contains 500 fill-in-the-blank evaluation instances. Within each subset, every evaluation instance corresponds to a unique structured educational objective. Specifically, two instances are regarded as different educational objectives if they differ in any component of $(G, K, D, S)$, including the grade level, knowledge concept set, difficulty level, or core competency set. Multiple-choice questions require plausible distractors and option-level consistency, whereas fill-in-the-blank questions place greater emphasis on precise formulation and answer uniqueness. Therefore, evaluating both subsets provides a more comprehensive assessment of EduAgentQG under different mathematics question formats.

The dataset serves two main roles in our experiments. First, for each evaluation instance, the structured educational objective $(G, K, D, S)$ is used as the input condition for question generation. EduAgentQG and all baseline methods receive only this educational objective as input. The corresponding human-written question is not provided to the Writer agent or to any baseline generation method. Instead, it is reserved as the gold-standard question for Win Rate evaluation, allowing us to compare generated questions with expert-designed questions under the same educational objective.

Second, the larger question bank is used by the Educator agent to retrieve reference questions related to the target grade level, knowledge concepts, difficulty level, and core competencies. These reference questions serve as anchors for evaluating difficulty alignment and core competency alignment, rather than as examples for question imitation. They are accessible only to the Educator agent during evaluation and are not exposed to the Writer agent during generation. To avoid information leakage, the gold-standard question corresponding to the current evaluation instance is excluded from reference retrieval.
\subsection{Experimental Setup}

\begin{table}[t]
\centering
\caption{Overview of the constructed mathematics question dataset and evaluation subsets.}
\label{tab:dataset_overview}
\resizebox{\linewidth}{!}{
\begin{tabular}{ll}
\toprule
\textbf{Item} & \textbf{Description} \\
\midrule
Total Questions & 10,273 \\
Grade Range & Grades 1--9 \\
Knowledge Concepts & 634 distinct concepts \\
Core Competencies & 16 categories \\
Difficulty Levels & Easy / Medium / Hard \\
\midrule
\textbf{Evaluation Subsets} & \\
MathChoice & 489 educational objectives (multiple-choice questions) \\
MathBlank & 500 educational objectives (fill-in-the-blank questions) \\
\bottomrule
\end{tabular}}
\end{table}

\textbf{Baselines.}
We compare EduAgentQG with four representative baselines:

\begin{itemize}
    \item \textbf{COT~\citep{wei2022chain}}: Uses chain-of-thought prompting to guide question generation through step-by-step reasoning without iterative refinement.
    \item\textbf{COT\_N~\citep{wang2022self}}: Samples $N$ reasoning paths for each case and selects the most consistent output by aggregating over the sampled reasoning paths.
    \item \textbf{ReAct~\citep{yao2023react}}: Interleaves reasoning and action, allowing the LLM to invoke external tools or retrieval when needed.
    \item \textbf{EQPR~\citep{cheng2025}}: Combines MCTS with LLMs and incorporates feedback into planning to refine question generation trajectories.
\end{itemize}

Following recent work on mathematics question generation~\cite{cheng2025}, we treat personalized mathematics question generation from structured educational objectives as an objective-conditioned reasoning and planning task. In this setting, a method is not only expected to produce fluent question text, but also to infer an appropriate mathematical context, construct a valid reasoning path, determine the answer, and satisfy the target grade level, knowledge concepts, difficulty level, and core competency constraints. Therefore, we select LLM-based reasoning and planning baselines that can be directly instantiated under the same structured educational objective and output requirements.

We do not include traditional template-based methods or earlier neural question generation models as direct baselines because their input assumptions and generation mechanisms differ substantially from our task setting. Template-based methods usually rely on predefined question schemas, fixed slots, and handcrafted transformation rules, which makes them suitable for highly regularized exercises but difficult to scale to objective-conditioned generation across 634 knowledge concepts, 16 core competency categories, three difficulty levels, and multiple question formats. Earlier neural question generation methods are typically designed for passage-based, answer-aware, or formula-conditioned generation, where the source text, target answer, equation, or problem structure is provided as input. In contrast, our task provides only the structured educational objective and requires the model to construct the question content and answer from that objective.

Accordingly, the selected baselines cover representative LLM-based paradigms that are compatible with this reasoning-oriented formulation. COT represents single-pass reasoning-based prompting, COT\_N represents multi-candidate reasoning-path sampling, ReAct represents reasoning with external actions or retrieval, and EQPR represents planning-based iterative refinement for mathematics question generation. This baseline selection follows the evaluation perspective of recent objective-based educational question generation studies while allowing direct comparison under the same personalized mathematics question generation protocol.

\begin{table*}[tb]
\centering
\caption{Performance comparison between EduAgentQG and baseline methods. Diversity metrics and Win Rate are reported as percentages, while all Objective Consistency dimensions are scored on a 0--10 scale. Bold numbers indicate the best performance in each column, while underlined numbers indicate the second-best performance.}
\label{tab:main_results}

\resizebox{\linewidth}{!}{
\begin{tabular}{l l c c c c c c c c c c c}
\toprule
& & \multicolumn{5}{c}{Diversity $\downarrow$}
& \multicolumn{5}{c}{Objective Consistency $\uparrow$}
& \multirow{2}{*}{Win Rate $\uparrow$} \\
\cmidrule(lr){3-7} \cmidrule(lr){8-12}
& & BLEU & METEOR & ROUGE-L & BERTScore & APCS
& Grade & Knowledge & Difficulty & Competency & Solvability & \\
\midrule

\multirow{17}{*}{MathChoice}
& & \multicolumn{11}{c}{Gemini-2.5-Flash} \\
& COT    & 26.08 & 56.75 & 50.61 & 81.30 & 58.29 & 9.3 & 9.0 & 8.2 & 7.9 & 8.8 & 63.53 \\
& COT\_N & 29.28 & 61.11 & 51.77 & 82.45 & 59.44 & 9.4 & 9.0 & 8.4 & 8.1 & 9.1 & 63.86 \\
& ReAct  & \underline{26.39} & \underline{56.49} & \underline{49.25} & \underline{80.65} & \underline{57.03} & \underline{9.6} & \underline{9.3} & \underline{8.4} & \underline{8.3} & \underline{9.3} & \underline{67.22} \\
& \textbf{EduAgentQG} & \textbf{16.40} & \textbf{45.78} & \textbf{40.13} & \textbf{74.56} & \textbf{48.05} & \textbf{9.7} & \textbf{9.5} & \textbf{8.6} & \textbf{8.8} & \textbf{9.5} & \textbf{71.12} \\
\cmidrule{2-13}

& & \multicolumn{11}{c}{GPT-4o-mini} \\
& COT    & 31.84 & 61.26 & 51.73 & 82.75 & \underline{55.72} & 9.3 & \textbf{9.1} & 7.8 & 8.1 & 8.8 & 24.97 \\
& COT\_N & 32.07 & \underline{50.96} & 51.83 & 81.62 & 59.78 & 9.4 & \textbf{9.1} & 7.7 & 8.0 & \underline{9.3} & \underline{28.62} \\
& ReAct  & 42.65 & 68.79 & 58.34 & 86.31 & 56.08 & 9.5 & \underline{8.9} & 7.8 & 8.2 & 9.0 & 26.02 \\
& EQPR   & \underline{31.22} & 55.63 & \underline{49.43} & \underline{79.21} & 58.24 & \underline{9.6} & 8.8 & \underline{7.9} & \underline{8.4} & 8.9 & 28.31 \\
& \textbf{EduAgentQG} & \textbf{17.34} & \textbf{48.28} & \textbf{46.63} & \textbf{75.22} & \textbf{50.60} & \textbf{9.8} & \textbf{9.1} & \textbf{8.0} & \textbf{8.6} & \textbf{9.6} & \textbf{32.20} \\
\cmidrule{2-13}

& & \multicolumn{11}{c}{Qwen2.5-72B-Instruct} \\
& COT    & 20.26 & 42.34 & 40.53 & 74.73 & \underline{51.60} & 9.2 & 9.1 & \underline{8.5} & 7.8 & 9.2 & 43.01 \\
& COT\_N & 27.83 & 58.88 & 49.10 & 81.51 & 57.43 & 9.4 & 9.2 & 8.4 & 8.0 & \underline{9.3} & 53.91 \\
& ReAct  & 36.12 & 64.39 & 52.74 & 84.39 & 59.02 & 9.5 & 9.3 & 8.4 & 8.0 & 9.0 & \underline{61.64} \\
& EQPR   & \underline{15.24} & \underline{51.94} & \underline{46.49} & \underline{71.83} & 52.11 & \underline{9.6} & \underline{9.4} & 8.3 & \underline{8.3} & 9.3 & 64.46 \\
& \textbf{EduAgentQG} & \textbf{12.75} & \textbf{39.07} & \textbf{38.32} & \textbf{70.98} & \textbf{46.55} & \textbf{9.7} & \textbf{9.5} & \textbf{8.7} & \textbf{8.4} & \textbf{9.6} & \textbf{66.33} \\

\midrule

\multirow{17}{*}{MathBlank}
& & \multicolumn{11}{c}{Gemini-2.5-Flash} \\
& COT    & 27.69 & 44.54 & 20.75 & 81.06 & 58.85 & 9.2 & 9.1 & 8.2 & 8.0 & 9.4 & 51.34 \\
& COT\_N & 27.85 & 44.37 & \underline{19.17} & 81.36 & 57.40 & 9.3 & 9.2 & 8.3 & 7.8 & \underline{9.5} & 48.66 \\
& ReAct  & \underline{26.89} & \underline{44.05} & 24.96 & \underline{80.02} & \underline{56.61} & \underline{9.5} & \underline{9.3} & \underline{8.5} & \underline{8.2} & \underline{9.5} & \underline{56.66} \\
& \textbf{EduAgentQG} & \textbf{22.23} & \textbf{38.86} & \textbf{18.26} & \textbf{76.39} & \textbf{50.02} & \textbf{9.6} & \textbf{9.6} & \textbf{8.7} & \textbf{8.6} & \textbf{9.7} & \textbf{61.41} \\
\cmidrule{2-13}

& & \multicolumn{11}{c}{GPT-4o-mini} \\
& COT    & 35.58 & 50.67 & 28.65 & 84.00 & 57.19 & 9.2 & 9.1 & 7.7 & 8.1 & 9.3 & 24.50 \\
& COT\_N & 31.97 & 46.55 & 28.62 & 82.28 & 55.59 & 9.3 & 9.1 & 7.7 & 8.0 & 9.3 & 26.00 \\
& ReAct  & 49.79 & 61.47 & 37.76 & 87.87 & 59.89 & 9.4 & 9.1 & \underline{7.8} & 8.1 & 9.1 & 25.55 \\
& EQPR   & \underline{25.19} & \underline{36.09} & \underline{27.70} & \underline{78.29} & \underline{52.05} & \underline{9.6} & \underline{9.2} & 7.7 & \underline{8.3} & 9.3 & \underline{40.45} \\
& \textbf{EduAgentQG} & \textbf{10.34} & \textbf{31.08} & \textbf{18.25} & \textbf{73.26} & \textbf{48.41} & \textbf{9.7} & \textbf{9.5} & \textbf{8.0} & \textbf{8.8} & \textbf{9.4} & \textbf{42.00} \\
\cmidrule{2-13}

& & \multicolumn{11}{c}{Qwen2.5-72B-Instruct} \\
& COT    & 22.34 & 39.41 & 27.13 & 79.25 & 53.77 & 9.3 & 9.0 & 8.3 & 8.0 & 9.1 & 43.85 \\
& COT\_N & 28.87 & 42.78 & 28.21 & 81.29 & 55.59 & 9.5 & 9.4 & \underline{8.5} & 8.1 & \underline{9.4} & 46.30 \\
& ReAct  & 47.41 & 61.37 & 46.94 & 88.11 & 59.47 & 9.4 & \underline{9.5} & 8.4 & 8.0 & 9.5 & 44.01 \\
& EQPR   & \underline{15.36} & \underline{33.64} & \underline{19.88} & \underline{74.95} & \underline{52.05} & \underline{9.6} & 9.2 & 8.3 & \underline{8.3} & 9.3 & \underline{49.76} \\
& \textbf{EduAgentQG} & \textbf{14.63} & \textbf{23.14} & \textbf{13.33} & \textbf{69.43} & \textbf{46.41} & \textbf{9.7} & \textbf{9.6} & \textbf{8.7} & \textbf{8.5} & \textbf{9.7} & \textbf{51.48} \\

\bottomrule
\end{tabular}
}
\end{table*}

\textbf{Evaluation Metrics.}
We evaluate generated questions from three perspectives: Diversity, Objective Consistency, and Win Rate. These metrics jointly measure whether the generated questions are varied, educationally aligned, and practically preferred.

\begin{itemize}
    \item \textbf{Diversity:} We use lexical and semantic similarity-based metrics to measure redundancy among questions generated under the same educational objective. Lower values indicate lower similarity and therefore higher diversity. BLEU~\citep{papineni2002bleu}, METEOR~\citep{banerjee2005meteor}, and ROUGE-L~\citep{lin2004looking} capture lexical or structural overlap, while BERTScore~\citep{zhang2019bertscore} and APCS~\citep{zhang2024improving} capture semantic similarity. Specifically, APCS denotes the average pairwise cosine similarity between generated question embeddings, where lower values indicate lower semantic redundancy and higher diversity. Using both lexical and semantic metrics helps distinguish surface-level variation from more meaningful differences in question formulation.
    
    \item \textbf{Objective Consistency:} We adopt an LLM-as-judge paradigm to evaluate whether each generated question satisfies the target educational objective. Objective Consistency is scored on a 0--10 scale, where higher scores indicate stronger alignment with the intended educational requirements. The evaluation considers five dimension-level criteria: grade-level alignment, knowledge-concept alignment, difficulty alignment, core-competency alignment, and solvability. These dimensions correspond to the main pedagogical and validity requirements of mathematics questions. Each dimension is evaluated according to the judge prompt provided in Appendix~A. This dimension-level rubric helps identify whether a lower score is caused by a mismatch in grade level, knowledge concept, difficulty, core competency, or mathematical validity, rather than relying only on an undifferentiated overall judgment.
    
    \item \textbf{Win Rate:} We compute the Win Rate relative to gold-standard questions. Each comparison records whether the generated question is preferred over the gold-standard question in terms of clarity, answer correctness, and pedagogical appropriateness. Unlike automatic similarity metrics, Win Rate provides an overall preference-based assessment of question quality, which is useful for evaluating whether generated questions are practically competitive with human-written questions.
\end{itemize}

\textbf{Implementation Details.}
The LLMs were configured with a temperature of 1.0 and top-$p$ sampling of 0.9. The backbone models include Gemini-2.5-Flash~\citep{Gemini25Report}, GPT-4o-mini~\citep{gpt-4o-mini}, and Qwen2.5-72B-Instruct~\citep{Qwen2TechnicalReport}. For automatic evaluation, DeepSeek-V3~\citep{deepseek2024} and GPT-4o~\citep{gpt-4o} were used as LLM judges, and their scores were averaged as the final evaluation results. To improve the transparency and reproducibility of the LLM-as-judge evaluation, we provide the complete prompts used for Objective Consistency and Win Rate evaluation in Appendix~\ref{app:judge_prompt}. The Objective Consistency prompt asks each judge to evaluate the generated question with respect to the given educational objective from five dimensions: grade-level alignment, knowledge concept alignment, difficulty alignment, core competency alignment, and solvability. The Win Rate prompt asks each judge to compare the model-generated question with the corresponding gold-standard question under the same educational objective, considering concept coverage, difficulty appropriateness, relevance to ability development, contribution to mathematical literacy, structural design, textual clarity, and problem accuracy.

Baseline methods follow their standard configurations with task-specific adaptations for personalized mathematics question generation. For comparability, all methods are evaluated under the same backbone model within each setting and receive the same structured educational objective, output format requirements, output-length constraints, and decoding settings. COT performs one-pass reasoning-based generation without retrieval, whereas COT\_N samples three reasoning paths under the same educational objective and selects the most consistent output. ReAct interleaves reasoning and action with retrieval when needed. Following the original setting of EQPR~\citep{cheng2025}, EQPR is configured with 4 iterations, a maximum depth of 3, and an exploration parameter of 2.5. EduAgentQG follows its planned multi-agent workflow, including planning, candidate generation, Solver and Educator evaluation, rewriting, and final checking.

Retrieval access follows each method's original inference protocol. When retrieval is used, ReAct and EduAgentQG access the same educational resources, including curriculum standards and knowledge graph information. Because the methods inherently differ in model calls, candidate selection, retrieval actions, and refinement or search opportunities, we do not claim identical inference budgets. Instead, we control the input objectives, backbone models, output requirements, output-length constraints, decoding settings, and shared retrieval resources where applicable. Pipeline-level computational cost is reported separately in Section~\ref{sec-cost}.

\subsection{Main Results}

We present the main comparison between EduAgentQG and baseline methods in Table~\ref{tab:main_results}. Due to the high computational cost of EQPR, it was not evaluated on the Gemini-2.5-Flash backbone. Across both MathChoice and MathBlank, EduAgentQG achieves the best overall performance across diversity, Objective Consistency, and Win Rate. This indicates that the proposed framework can improve question quality while also reducing redundancy among questions generated under the same educational objective.

On Gemini-2.5-Flash, EduAgentQG improves Win Rate by 3.90 percentage points on MathChoice and 4.75 percentage points on MathBlank compared with the strongest baseline, ReAct. It also consistently lowers BLEU, METEOR, ROUGE-L, BERTScore, and APCS. Since these metrics measure lexical or semantic similarity, the lower values suggest that EduAgentQG generates questions with more diverse formulations and assessment perspectives. This improvement is especially important for personalized education, where multiple questions under the same objective should not merely repeat the same structure with minor numerical changes.

Similar trends are observed on GPT-4o-mini and Qwen2.5-72B-Instruct. On GPT-4o-mini, EduAgentQG substantially improves Win Rate over all baselines on both subsets, while also achieving lower similarity-based diversity metrics. On Qwen2.5-72B-Instruct, although EQPR performs competitively in Win Rate, EduAgentQG still achieves the best overall results and shows stronger performance in Objective Consistency. These results suggest that the proposed framework is not tied to a specific backbone model, but can generalize across LLMs with different capabilities.

In terms of Objective Consistency, EduAgentQG consistently improves grade-level alignment, knowledge concept alignment, difficulty alignment, core competency alignment, and solvability. The gains in core competency alignment and solvability are particularly meaningful, because these dimensions are often difficult to control through direct prompting alone. The results suggest that the Solver and Educator agents provide complementary verification: the Solver focuses on logical validity and answer correctness, while the Educator checks whether the question satisfies pedagogical requirements. This division of roles helps prevent questions from passing evaluation merely because they are fluent or superficially relevant.

Overall, the main results demonstrate that structured planning, multi-agent evaluation, and iterative refinement jointly contribute to higher-quality question generation. Compared with single-agent prompting methods, EduAgentQG provides stronger control over both diversity and educational alignment. Compared with EQPR, EduAgentQG achieves a more balanced improvement across multiple evaluation dimensions, rather than optimizing only overall preference or correctness.

\subsection{Human Evaluation Results}

\begin{table}[t]
\centering
\caption{Human evaluation results and agreement statistics. Objective Consistency is scored on a 0--10 scale, Win Rate is reported as a percentage, and agreement is measured using weighted kappa for Objective Consistency and kappa for Win Rate.}
\label{tab:human_eval}
\small
\setlength{\tabcolsep}{3pt}
\renewcommand{\arraystretch}{1.15}
\begin{tabular}{@{}p{0.40\columnwidth}!{\vrule width 0.8pt}>{\centering\arraybackslash}p{0.33\columnwidth}>{\centering\arraybackslash}p{0.20\columnwidth}@{}}
\toprule
 & \textbf{Objective Consistency} & \textbf{Win Rate} \\
\midrule
COT                 & 8.09 & 39.22 \\
COT\_N              & 8.18 & 33.12 \\
ReAct               & 8.24 & 41.46 \\
EQPR                & 7.97 & 47.37 \\
\textbf{EduAgentQG} & \textbf{8.42} & \textbf{53.34} \\
\midrule
Human--Human        & $\kappa_w = 0.72$ & $\kappa = 0.71$ \\
Human--DeepSeek-V3  & $\kappa_w = 0.68$ & $\kappa = 0.63$ \\
Human--GPT-4o       & $\kappa_w = 0.70$ & $\kappa = 0.66$ \\
Human--LLM ensemble & $\kappa_w = 0.75$ & $\kappa = 0.70$ \\
\bottomrule
\end{tabular}
\end{table}

\begin{table*}[t]
\centering
\small
\caption{Representative disagreement cases between human evaluators and LLM judges.}
\label{tab:disagreement_cases}
\begin{tabularx}{\textwidth}{p{0.36\textwidth} p{0.29\textwidth} p{0.29\textwidth}}
\toprule
\textbf{Evaluation case} & \textbf{Human judgment} & \textbf{LLM judgment} \\
\midrule

\textbf{Objective Consistency case.}

\textbf{Target objective:} Grade 6; percentage; hard difficulty; reasoning.

\textbf{Generated question:} A schoolbag is sold at a 20\% discount. After the discount, the price is 48 yuan. What was the original price of the schoolbag?

\textbf{Answer:} 60 yuan.
&
\textbf{Score: 6/10.} The question has complete conditions and a unique answer. It correctly examines the relationship between the discounted price and the original price in a percentage problem. However, the solution mainly involves one reverse calculation, $48 \div 0.8 = 60$, and the reasoning process is relatively shallow. Therefore, the question is closer to a routine percentage calculation problem and does not fully match the target hard difficulty or reasoning.
&
\textbf{Score: 8/10.} The question is clearly related to percentage, has sufficient conditions, and is easy to understand. The answer is unique and the solution requires students to infer the original price from the discounted price. The LLM judges regarded this reverse calculation process as evidence of reasoning and considered the question generally consistent with the target objective. \\

\midrule

\textbf{Win Rate case.}

\textbf{Target objective:} Grade 5; area of trapezoids; medium difficulty; geometric intuition.

\textbf{Generated question:} A trapezoid has an upper base of 6 cm, a lower base of 10 cm, and a height of 4 cm. What is its area?

\textbf{Answer:} 32 cm$^2$.

\textbf{Gold-standard question:} A trapezoid can be divided into a rectangle and two congruent right triangles. The rectangle is 8 cm wide and 5 cm high, and each right triangle has legs of 2 cm and 5 cm. What is the area of the trapezoid?

\textbf{Answer:} 50 cm$^2$.
&
\textbf{Preference: Gold-standard question.} The gold-standard question examines not only the area of a trapezoid but also the structural relationship between a trapezoid, a rectangle, and triangles. Students need to understand the geometric decomposition and combine the areas of different shapes. In comparison, the generated question directly provides the upper base, lower base, and height, so students only need to substitute values into the trapezoid area formula. Therefore, the gold-standard question better reflects the target geometric intuition and medium difficulty.
&
\textbf{Preference: Generated question.} The generated question is concise, unambiguous, and directly aligned with the knowledge concept of trapezoid area. The conditions are clearly given, the calculation goal is explicit, and the answer is unique. In comparison, the gold-standard question contains more conditions and a more complex description, which may increase the reading and understanding burden. Therefore, the LLM judges preferred the generated question. \\

\bottomrule
\end{tabularx}
\end{table*}

\begin{figure*}[htpb]
    \begin{adjustwidth}{0pt}{0pt}
        \vspace{-6mm}

        \begin{subfigure}[t]{0.94\textwidth}
            \includegraphics[width=\textwidth]{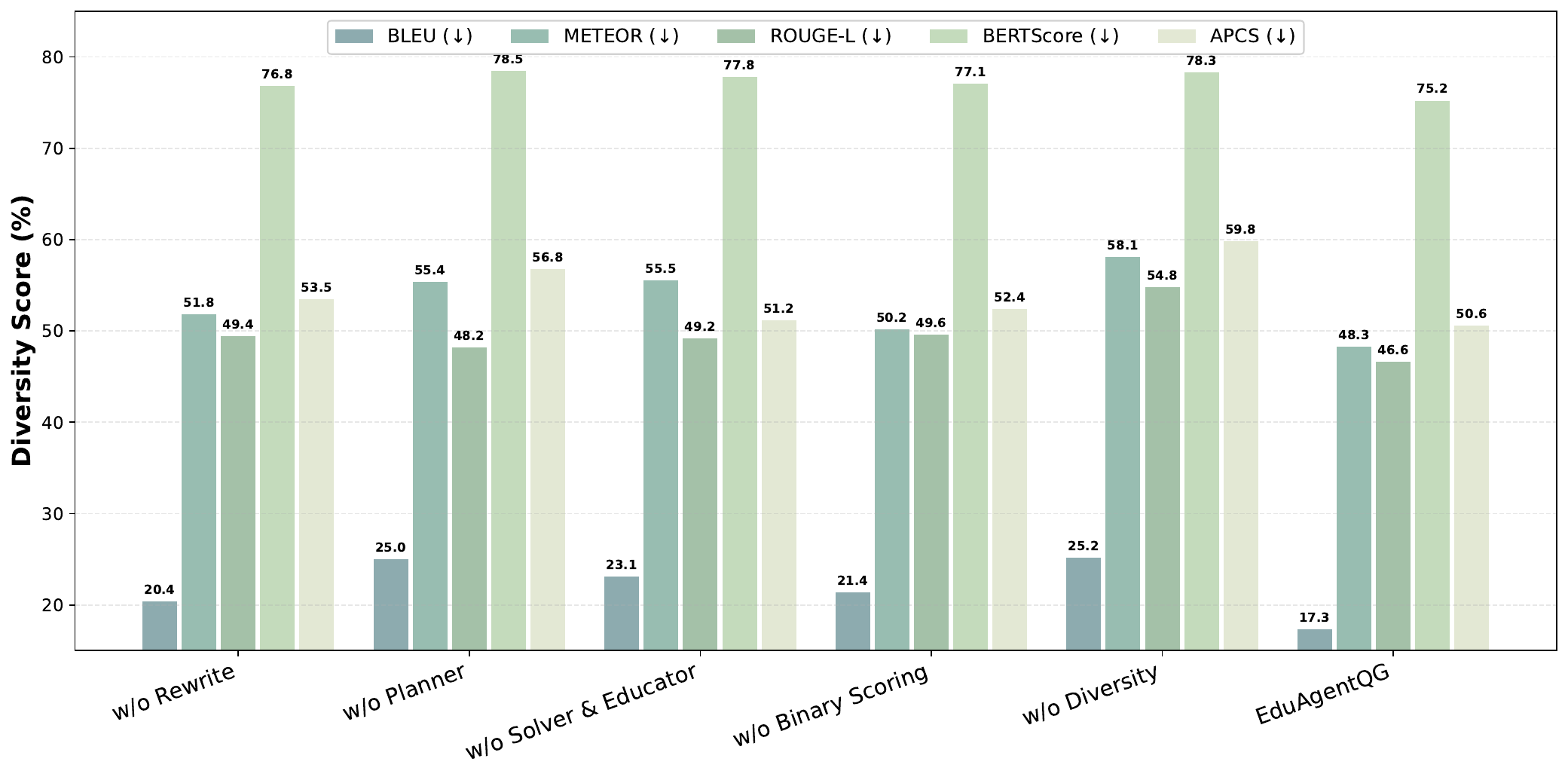}
            \caption{Similarity-based diversity metrics ($\downarrow$)}
            \label{fig:diversity}
        \end{subfigure}

        \vspace{1em}

        \begin{subfigure}[t]{\textwidth}
            \includegraphics[width=\textwidth]{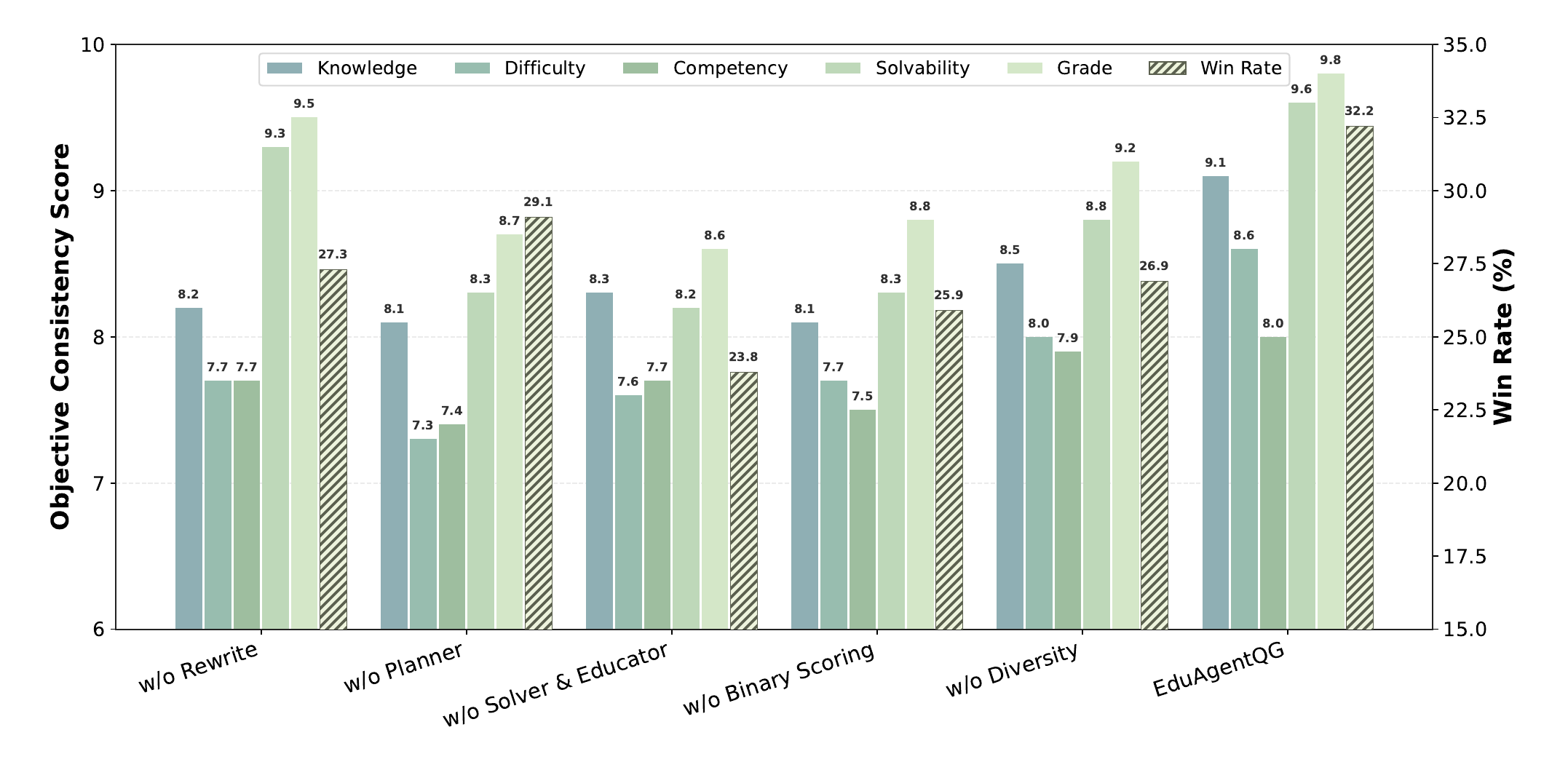}
            \caption{Objective Consistency and Win Rate comparison ($\uparrow$)}
            \label{fig:ablation2}
        \end{subfigure}
    \end{adjustwidth}

    \caption{Ablation results of EduAgentQG on the MathChoice dataset using GPT-4o-mini. (a) Similarity-based diversity metrics, where lower values indicate lower similarity and thus higher diversity. (b) Objective Consistency and Win Rate; the left y-axis denotes Objective Consistency scores (0--10), and the right y-axis denotes Win Rate (0--100\%).}
    \label{fig:ablation}
\end{figure*}

To further validate the practical utility of EduAgentQG, we conducted a human evaluation study. Three researchers with expertise in mathematics education were recruited as evaluators. From each evaluation subset, 50 questions were randomly sampled, resulting in 100 evaluation instances. All questions were anonymized and presented in random order to reduce potential bias, so the evaluators did not know which method produced each question.

The evaluation focuses on Objective Consistency and Win Rate. Objective Consistency is scored on a 0--10 scale according to the alignment between the generated question and the target educational objective. Win Rate is computed as the percentage of cases in which a generated question is preferred over the corresponding gold-standard question. For Objective Consistency, the final score is averaged across the three evaluators; for Win Rate, the final decision is determined by majority voting. This design allows us to assess both dimension-level alignment and overall human preference.

To assess reliability, we report agreement statistics in Table~\ref{tab:human_eval}. For the three human evaluators, we report the average pairwise Cohen's weighted kappa for Objective Consistency and the average pairwise Cohen's kappa for Win Rate. The human agreement reaches $\kappa_w=0.72$ for Objective Consistency and $\kappa=0.71$ for Win Rate, indicating substantial consistency among evaluators. This suggests that the evaluation criteria are sufficiently clear for human experts to apply consistently.

We also compare human judgments with automatic judgments from DeepSeek-V3 and GPT-4o. The LLM ensemble achieves the highest agreement with humans among the automatic judgment settings, with $\kappa_w=0.75$ for Objective Consistency and $\kappa=0.70$ for Win Rate. This result supports the use of multiple LLM judges for automatic evaluation, as ensembling can reduce model-specific bias and better approximate human expert judgment.

To better understand the remaining differences between automatic evaluation and human evaluation, we further analyzed representative cases where the LLM ensemble disagreed with the majority human judgment. Table~\ref{tab:disagreement_cases} presents two typical disagreement cases, covering Objective Consistency evaluation and Win Rate evaluation, respectively.

As shown in Table~\ref{tab:disagreement_cases}, disagreements mainly arise when difficulty or core competency alignment requires subtle pedagogical judgment. LLM judges tend to emphasize formal correctness, clarity, and knowledge-concept relevance, whereas human evaluators place greater weight on whether a question reflects the intended cognitive demand and competency. These cases show that explicit LLM-as-judge prompts support scalable evaluation, while human evaluation remains important for validating nuanced pedagogical judgments.

For human evaluation, all methods were implemented using Qwen2.5-72B-Instruct to ensure a fair comparison under the same model capacity. As shown in Table~\ref{tab:human_eval}, EduAgentQG achieves the highest Objective Consistency score of 8.42 and the highest human-evaluated Win Rate of 53.34\%. Compared with the baselines, EduAgentQG is more frequently preferred by human evaluators, indicating that its improvements are not limited to automatic metrics.

Qualitative feedback further supports these results. Evaluators noted that questions generated by EduAgentQG generally exhibit clearer logical structures, fewer ambiguities, and stronger alignment with the intended educational objectives. In particular, EduAgentQG is better at maintaining appropriate difficulty while reflecting the target knowledge concept and core competency. This confirms that the proposed multi-agent framework improves not only measurable performance, but also the practical pedagogical quality of generated questions.

\begin{figure*}
    \centering

    \begin{subfigure}[t]{0.48\textwidth}
        \centering
        \includegraphics[width=\linewidth]{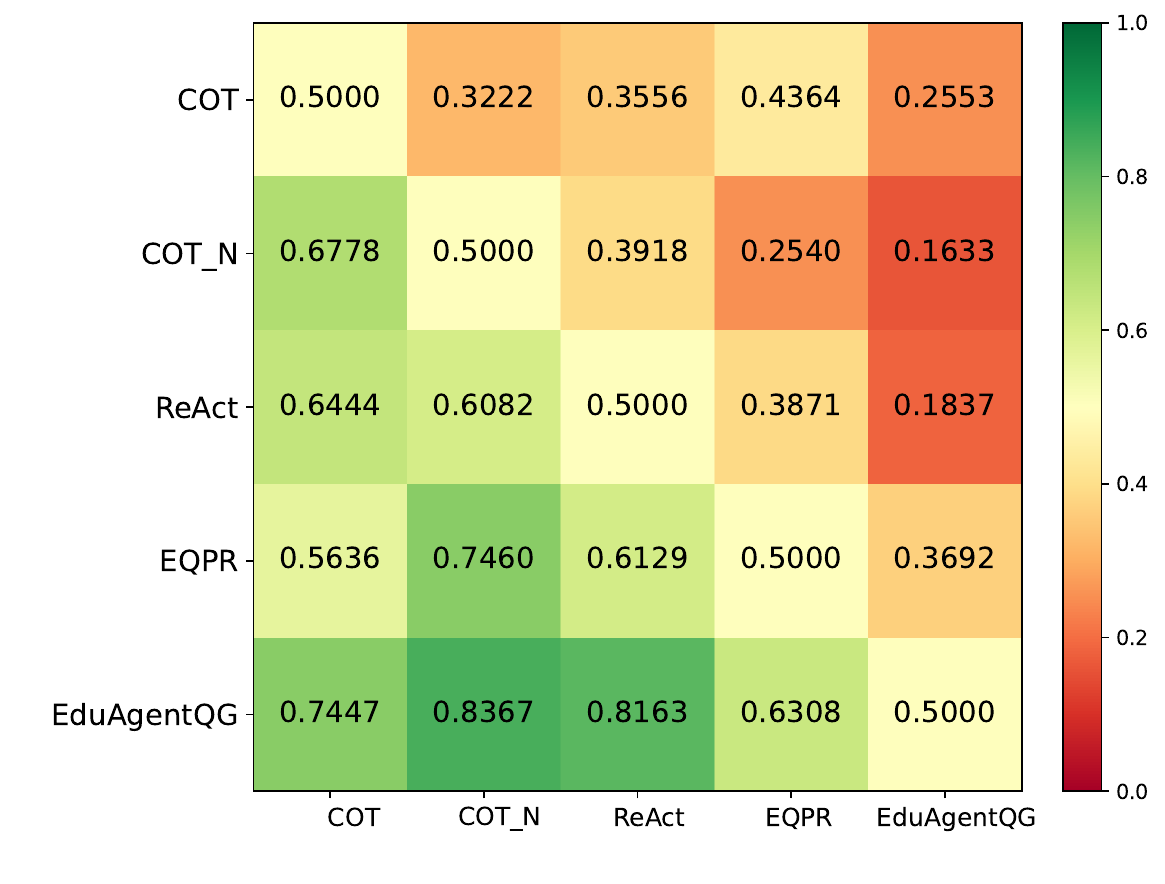}
        \caption{MathChoice dataset}
        \label{fig:pairwise_heatmap_mathchoice}
    \end{subfigure}
    \hfill
    \begin{subfigure}[t]{0.48\textwidth}
        \centering
        \includegraphics[width=\linewidth]{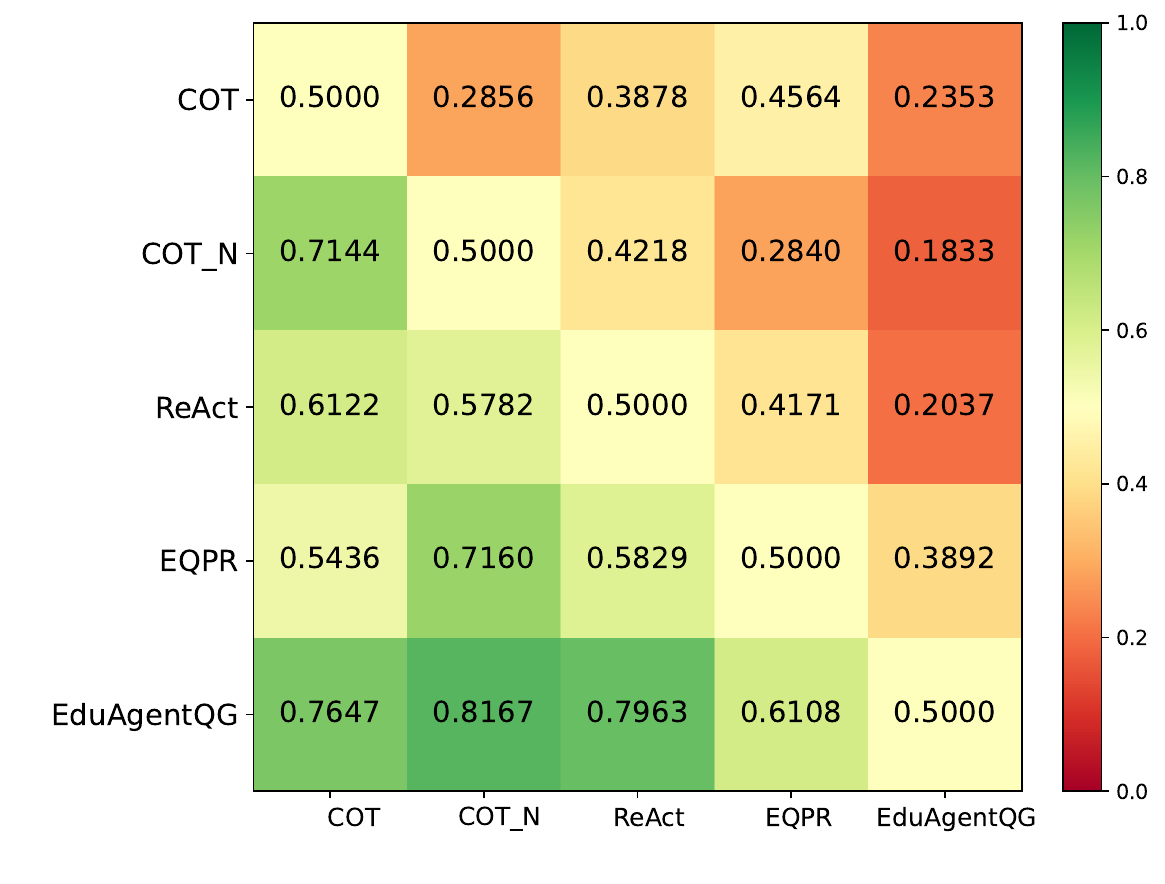}
        \caption{MathBlank dataset}
        \label{fig:pairwise_heatmap_mathblank}
    \end{subfigure}

    \caption{Heatmaps of pairwise Win Rate comparisons among different methods on the MathChoice and MathBlank datasets using GPT-4o-mini. Each cell denotes the Win Rate of the row method over the column method. Values greater than 0.5 indicate that the row method is more frequently preferred.}
    \label{fig:pairwise_heatmap}
\end{figure*}

\begin{figure*}
    \centering
    \includegraphics[width=0.9\linewidth]{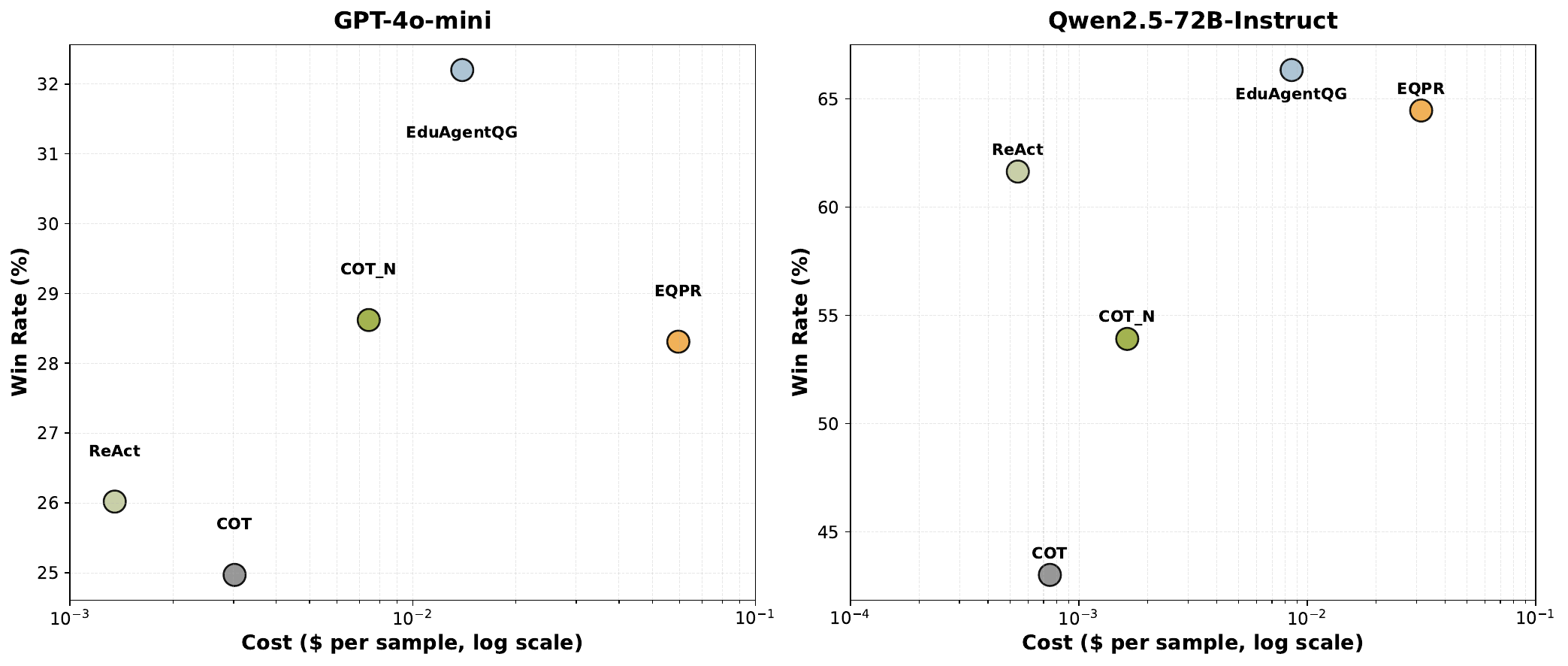}
    \caption{Cost--performance trade-off comparison of different methods. The x-axis denotes the average cost per sample (USD, log scale), and the y-axis denotes the Win Rate against gold-standard questions.}
    \label{fig:cost}
\end{figure*}

\subsection{Ablation Results}

To evaluate the contribution of each component, we conduct an ablation study on MathChoice using GPT-4o-mini. We remove the rewriting module (w/o Rewrite), the planner (w/o Planner), the Solver and Educator agents (w/o Solver \& Educator), the binary scoring mechanism (w/o Binary Scoring), and the diversity constraint (w/o Diversity). The complete EduAgentQG framework serves as the full model.

As shown in Figure~\ref{fig:ablation}, removing the rewriting module reduces Win Rate, indicating that iterative optimization is important for improving the final output quality. Without rewriting, the system cannot effectively incorporate feedback from the evaluation agents, so initially flawed candidates are more likely to remain unresolved. This confirms the importance of closed-loop refinement in the generation process.

Removing the planner leads to increased similarity-based diversity metrics and lower Objective Consistency. This result shows that structured planning is important for both diversity and educational alignment. Without explicit generation directions, the Writer tends to rely more heavily on the inherent randomness of the LLM, which may produce surface-level variation but is less effective at generating meaningfully different question structures.

The removal of the Solver and Educator agents causes the largest drop in Win Rate. This demonstrates that evaluation agents play a central role in filtering and improving candidate questions. The Solver mainly ensures logical soundness, solvability, and answer correctness, while the Educator focuses on grade-level alignment, knowledge concept alignment, difficulty alignment, and core competency alignment. Removing both agents weakens the framework's ability to identify and correct low-quality candidates.

The binary scoring mechanism also contributes to robust candidate selection. Without binary judgments, the system relies more on aggregated or ranking-based feedback, which may allow questions with specific dimension-level failures to pass. In contrast, binary pass/fail signals provide stricter control over essential requirements such as solvability and objective alignment.

Finally, removing the diversity constraint increases several similarity-based metrics and reduces competency alignment. This suggests that diversity control does not merely produce surface-level variation; it also encourages the model to explore different reasoning paths and assessment forms. Overall, the full model achieves the best performance, showing that planning, rewriting, evaluation, binary scoring, and diversity control work together to improve question quality, alignment, and diversity.

\subsection{Pairwise Win Rate Evaluation Results}

\begin{figure*}
    \centering
    \vspace{-2.1em}
    \includegraphics[width=\linewidth]{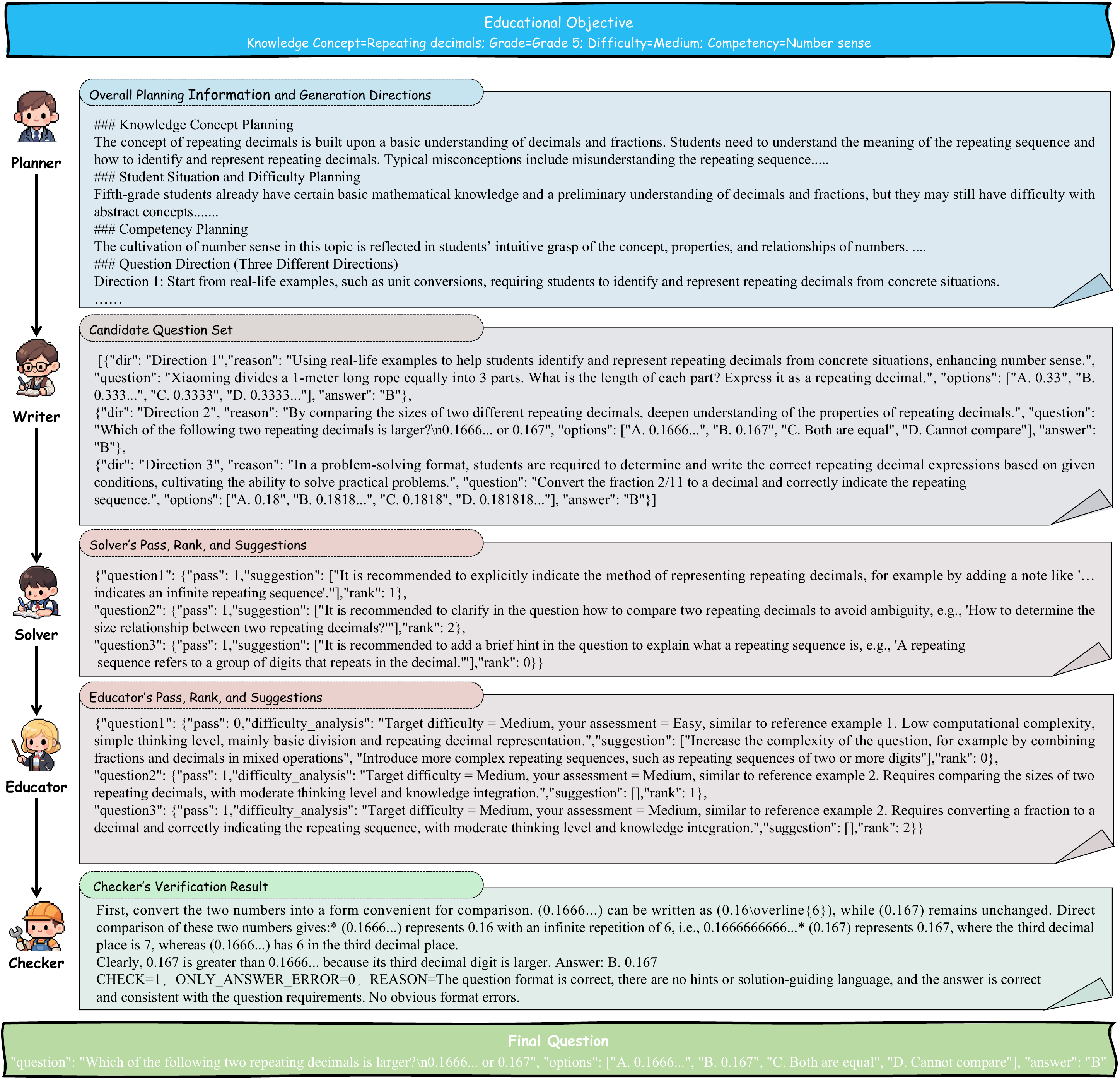}
    \caption{A case study illustrating the EduAgentQG generation workflow on a question about repeating decimals. The Planner defines three pedagogical directions: identifying repeating decimals in real-life contexts, comparing the magnitude of different repeating decimals, and converting fractions to repeating decimals. The Writer generates candidate questions based on these directions. The Solver checks logical correctness and provides revision suggestions. The Educator evaluates difficulty alignment and pedagogical suitability. The Checker verifies solvability and ambiguity to produce the final question.}
    \label{fig:case2}
\end{figure*}

To provide a direct comparison among methods, we visualize pairwise Win Rates using heatmaps, as shown in Figure~\ref{fig:pairwise_heatmap}. Each cell denotes the Win Rate of the row method over the column method, and values above $0.5$ indicate that the row method is more frequently preferred. This visualization complements the main results by showing not only each method's performance against gold-standard questions, but also the relative preference between generated outputs from different methods.

EduAgentQG consistently obtains values above $0.5$ against all baselines on both datasets, showing that it is more frequently preferred in direct comparisons. This pattern is stable across MathChoice and MathBlank, indicating that EduAgentQG's advantage is not limited to a single question format. In contrast, the baseline methods show more mixed results. COT and COT\_N are generally less competitive, suggesting that direct prompting and self-consistency-based aggregation are insufficient for stable objective-aware question generation.

ReAct and EQPR improve over COT-based methods in several cases, but they are still surpassed by EduAgentQG. ReAct benefits from reasoning--action interleaving, while EQPR benefits from iterative search and feedback. However, neither method explicitly decomposes educational objectives into multiple generation directions or performs strict dimension-wise evaluation. The pairwise results therefore further confirm the advantage of combining structured planning, multi-agent evaluation, and closed-loop refinement.

\subsection{Cost--Performance Trade-off Analysis}
\label{sec-cost}
\begin{figure*}
    \centering
    \includegraphics[width=\linewidth]{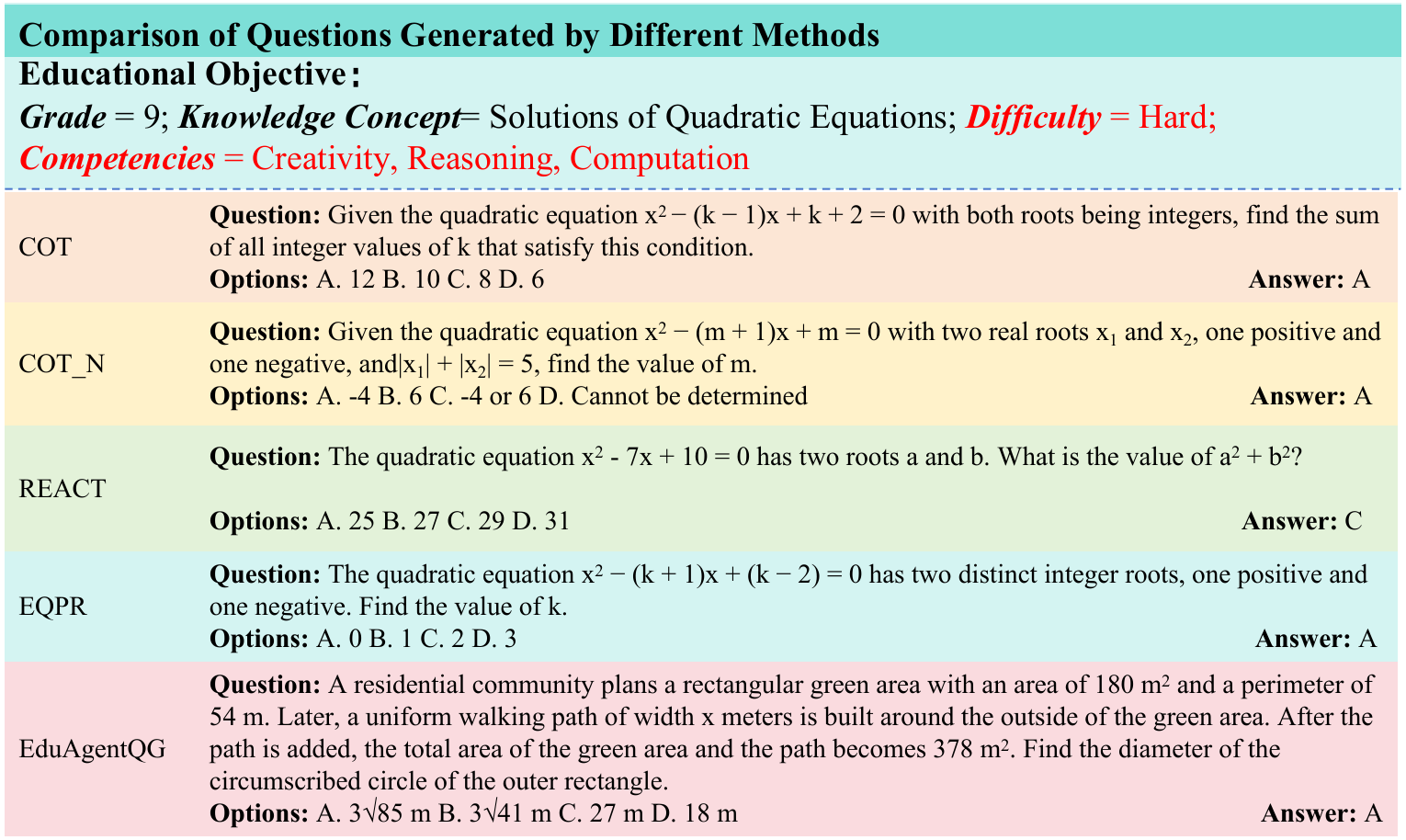}
    \caption{A comparative case study of questions generated by different methods under the same educational objective. Baseline methods satisfy basic solvability but show limitations in difficulty alignment, reasoning depth, or core competency coverage, whereas EduAgentQG better aligns with the target objective through a scenario-based and multi-step reasoning design.}
    \label{fig:case_study_english_final}
\end{figure*}

\begin{figure*}
    \centering
    \includegraphics[width=\linewidth]{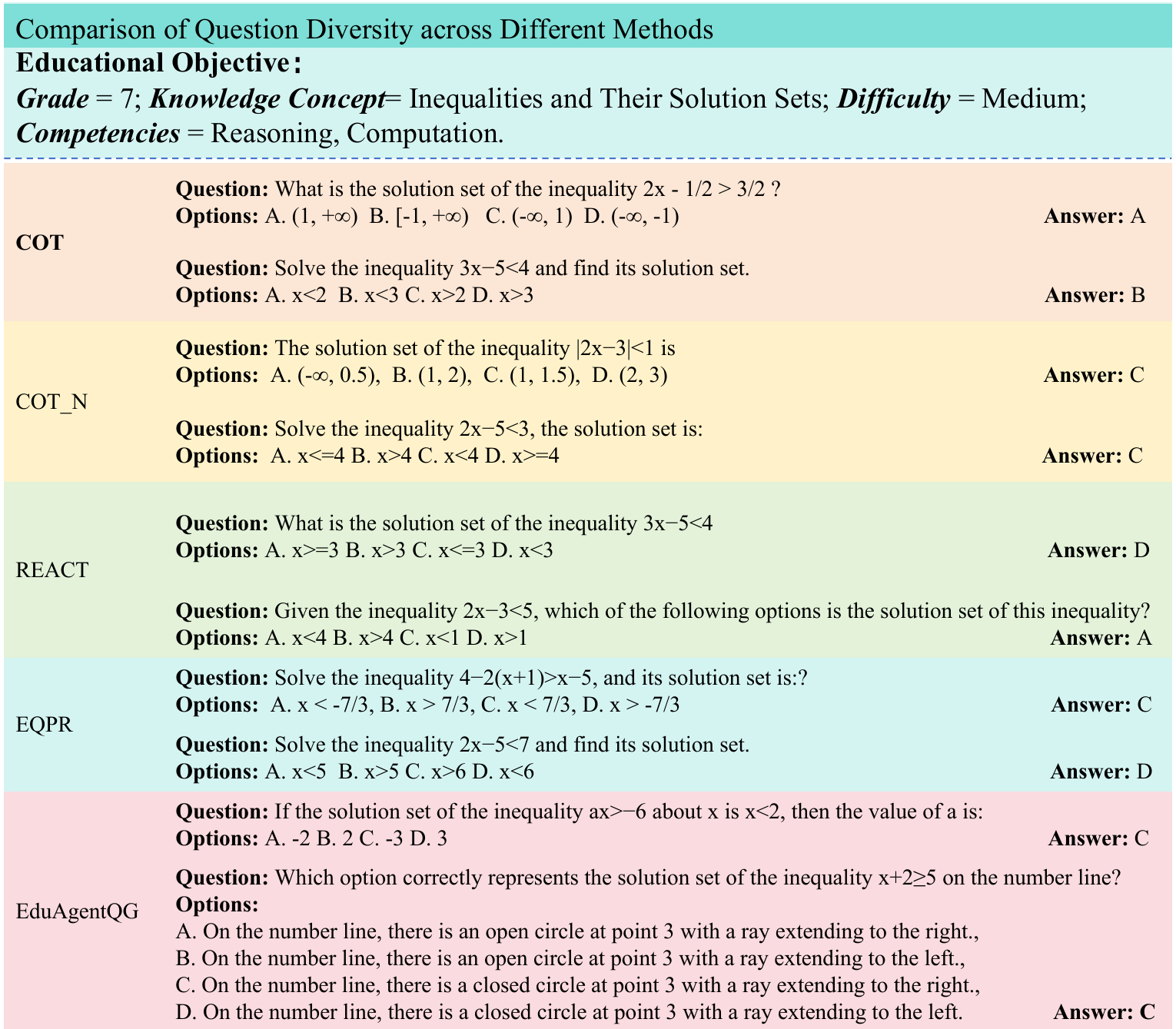}
    \caption{A comparative case study of question diversity across different methods under the same educational objective. Compared with baselines that mainly vary coefficients, inequality symbols, or interval endpoints, EduAgentQG produces structurally more diverse questions involving parameter reasoning and number-line representation.}
    \label{fig:case_study_diversity}
\end{figure*}

To assess cost-efficiency, we compare EduAgentQG with COT, COT\_N, ReAct, and EQPR using GPT-4o-mini and Qwen2.5-72B-Instruct. All methods are evaluated on the same query set, and the cost of each method is calculated by summing the input and output token costs incurred during its actual inference process. For EduAgentQG, this pipeline-level cost includes all agent calls, including planning, candidate generation, Solver and Educator evaluation, rewriting, and final checking. For the baseline methods, the reported cost includes their corresponding prompting, sampling, retrieval, or search steps according to their inference procedures. 

As shown in Figure~\ref{fig:cost}, EduAgentQG achieves the best cost--performance trade-off across both models. With GPT-4o-mini, EduAgentQG obtains a Win Rate of 32.2\% at a cost of 0.0139 USD per sample, outperforming EQPR, which achieves 28.3\% at 0.0596 USD per sample. This means that EduAgentQG achieves better performance while requiring substantially lower cost. The advantage mainly comes from its structured workflow, which reduces unnecessary exploration and focuses generation and refinement on targeted educational requirements.

With Qwen2.5-72B-Instruct, EduAgentQG reaches a Win Rate of 66.3\% at 0.0085 USD per sample, exceeding EQPR by 1.9 percentage points while requiring substantially lower cost. Although the absolute Win Rate is higher with Qwen2.5-72B-Instruct due to its stronger reasoning capability, the relative advantage of EduAgentQG remains consistent. This suggests that the framework can benefit from stronger backbone models while maintaining favorable cost-efficiency.

These results show that EduAgentQG improves question generation quality without relying on excessive computational expenditure. Compared with methods that depend on extensive search or repeated sampling, EduAgentQG uses explicit planning and targeted evaluation to improve generation efficiency. Its cost-efficiency advantage across different backbones suggests that the proposed framework is practical for scalable LLM-based educational applications.

\subsection{Case Study}

To illustrate how EduAgentQG operates in practice, we present three qualitative analyses: workflow analysis, generated question comparison, and question diversity comparison.

\subsubsection{Workflow Analysis}

Figure~\ref{fig:case2} presents a representative workflow case on repeating decimals. The Planner first analyzes the educational objective, identifies prerequisite knowledge and common misconceptions, and defines three generation directions: real-life identification, magnitude comparison, and fraction-to-decimal conversion. These directions provide explicit guidance for the Writer, ensuring that candidate questions are not generated from a single narrow perspective.

Based on these directions, the Writer generates candidate multiple-choice questions. Each candidate follows the same educational objective but differs in context or reasoning form. For example, one question uses a real-life division scenario, another requires comparison between decimals, and another asks students to convert a fraction into a repeating decimal. This shows how the planning stage supports controllable diversity before evaluation begins.

The Solver then checks logical correctness, clarity, and solvability, while the Educator evaluates whether each candidate matches the target difficulty and competency requirement. Their feedback is used to revise or rank candidates. In this case, the Educator identifies that one candidate is too easy for the target difficulty, while other candidates better match the medium-level requirement. This demonstrates the role of fine-grained evaluation in preventing superficially correct but pedagogically mismatched questions from being selected.

Finally, the Checker verifies the selected question for answer correctness and ambiguity. This case shows how EduAgentQG transforms an educational objective into a valid and difficulty-aligned question through coordinated planning, generation, evaluation, and verification.

\subsubsection{Generated Question Comparison}

Figure~\ref{fig:case_study_english_final} compares questions generated under the same educational objective: Grade 9 solutions of quadratic equations, hard difficulty, and core competencies of Creativity, Reasoning, and Computation. COT and COT\_N generate solvable questions, but their difficulty and reasoning depth are relatively limited. Their questions mainly focus on algebraic manipulation and do not fully reflect the intended competency requirements.

ReAct introduces more reasoning steps, but its assessment of integrated competencies remains insufficient. The generated question is mathematically valid, yet it mainly tests direct use of root relationships and lacks a richer problem context. EQPR improves condition completeness, but its problem structure is still relatively narrow and does not sufficiently emphasize creativity or higher-order reasoning.

In contrast, EduAgentQG generates a scenario-based problem that combines quadratic-equation reasoning with geometric constraints. The question requires students to integrate area, perimeter, and circumscribed-circle relationships, thereby better matching the target difficulty and competency requirements. Compared with the baselines, this question requires multi-step reasoning rather than direct formula substitution. This comparison illustrates the advantage of EduAgentQG in aligning generated questions with multi-dimensional educational objectives.

\subsubsection{Question Diversity Comparison}

To examine diversity under the same educational objective, we conduct a qualitative comparison on MathChoice using GPT-4o-mini. As shown in Figure~\ref{fig:case_study_diversity}, all methods generate questions for Grade 7 inequalities and their solution sets, with medium difficulty and core competencies of Reasoning and Computation. Since the same base LLM and the same educational objective are used, the comparison focuses on how different prompting or framework designs affect question diversity.

The baseline methods mainly produce direct inequality-solving questions. Their variations are mostly limited to coefficients, inequality symbols, or interval endpoints, while the underlying structure remains similar. In other words, although the surface forms differ, the cognitive operation required from students remains largely unchanged: solve a single linear inequality and select the correct interval. This suggests that general prompting strategies tend to keep the same base LLM within a narrow formulation space.

EduAgentQG produces more structurally diverse questions. Besides direct computation, it introduces parameter-based reasoning and representation-oriented tasks such as interpreting solution sets on a number line. These questions remain aligned with the same knowledge concept and competency requirements, but assess the concept through different cognitive and representational forms. This case indicates that EduAgentQG improves diversity at the structural level rather than relying only on surface-level lexical or numerical variation.

\section{Discussion}

\subsection{Main Findings}

The experimental results suggest that the central challenge in personalized question generation is not only to generate mathematically correct questions, but also to balance Objective Consistency and controllable diversity under explicit educational objectives. Across MathChoice and MathBlank, EduAgentQG consistently improves diversity, Objective Consistency, and Win Rate over COT, COT\_N, ReAct, and EQPR. Similar patterns across different backbone models show that these improvements are not limited to a specific model setting. These results directly respond to the two challenges discussed in the Introduction. First, questions should not be accepted because they satisfy an overall quality criterion; they must also meet specific educational requirements, including grade-level alignment, knowledge-concept alignment, difficulty alignment, core-competency alignment, and solvability. The improvements in Objective Consistency and Win Rate show that fine-grained evaluation helps maintain these requirements during generation and refinement. Second, questions generated under the same educational objective should not repeat similar semantic content or assessment perspectives. The lower similarity-based diversity metrics show that decomposing an educational objective into multiple generation directions can reduce redundancy while preserving educational alignment. The ablation results support this interpretation. Removing the Planner weakens both diversity and Objective Consistency, while removing the Solver and Educator causes the largest drop in Win Rate. Removing the rewriting module reduces Win Rate, indicating that evaluation feedback needs to be incorporated into refinement rather than used only for candidate assessment. Overall, these results show that structured planning, fine-grained evaluation, and feedback-driven rewriting work together to improve question diversity, educational alignment, and overall generation quality.

\subsection{Comparison with Existing Studies}

Compared with existing studies, EduAgentQG mainly differs in two aspects: question diversity and evaluation reliability. First, in diversity control, existing LLM-based generation methods often rely on model randomness or implicit reasoning patterns to obtain diverse outputs~\citep{karbasi2025,cheng2025}. When generation is conditioned on the same educational objective, such reliance may lead to questions with similar semantic content and assessment perspectives. EduAgentQG addresses this limitation through generation-direction-based diversity control. Specifically, the Planner transforms each educational objective into structured planning information and multiple generation directions, which guide the Writer to generate questions with different contexts, presentation styles, and reasoning paths. As shown in Figure~\ref{fig:ablation}, removing the Planner weakens both diversity and Objective Consistency, indicating that structured planning contributes to improving question diversity while maintaining educational alignment. Figure~\ref{fig:case_study_diversity} further shows that EduAgentQG produces more structurally diverse questions, whereas the baselines mainly vary numerical or symbolic elements. Second, in question evaluation, existing methods often rely on coarse-grained, aggregated, or implicitly represented feedback~\citep{zhou2023,karbasi2025,cheng2025,wang2024}, which may obscure deficiencies in specific educational dimensions. EduAgentQG addresses this limitation through fine-grained objective-aware evaluation. Specifically, the Solver evaluates logical soundness, solvability, and ambiguity, while the Educator evaluates grade-level, knowledge-concept, difficulty, and core-competency alignment. Candidates that fail either evaluation are revised by the Writer using the corresponding feedback. As shown in Figure~\ref{fig:ablation}, removing the Solver and Educator causes the largest drop in Win Rate, providing further evidence for the importance of dimension-wise evaluation. Overall, EduAgentQG differs from existing methods by reducing reliance on model randomness through generation-direction-based diversity control and by replacing coarse-grained or aggregated feedback with fine-grained evaluation across logical and educational dimensions.

\subsection{Theoretical Implications}

From a theoretical perspective, this study frames personalized question generation as an objective-aware educational generation task rather than a general text generation task. Previous question generation methods often emphasize fluency, correctness, or overall quality, but educational questions require a stronger connection to predefined instructional constraints. A generated question may be fluent and solvable, yet still fail to match the intended knowledge concept, difficulty level, grade level, or core competency. Therefore, relying only on aggregated feedback or overall scoring is insufficient for explaining whether a question truly satisfies an educational objective. EduAgentQG addresses this issue by connecting educational objective decomposition, controllable generation, and dimension-wise evaluation in a unified framework. Compared with single-agent prompting methods, EduAgentQG does not require the same model to simultaneously perform planning, generation, reasoning, and self-evaluation. Compared with existing multi-agent or feedback-based methods, it introduces explicit generation directions to guide diversity and uses the Solver and Educator to separately verify logical soundness and educational alignment. This design makes the quality control process more interpretable and provides a more structured way to model personalized educational question generation.

\subsection{Practical Implications}

From a practical perspective, EduAgentQG can provide controllable support for classroom teaching, adaptive assessment, and question bank construction. In real teaching scenarios, educators often need to design multiple questions around the same educational objective. However, these questions should not be produced through simple numerical replacement or minor wording changes. Instead, they should vary in context, phrasing, reasoning paths, and assessment perspectives while still maintaining the intended difficulty level and grade-level appropriateness. EduAgentQG supports this need through its multi-agent workflow. The Planner produces structured planning information and generation directions, the Writer generates and revises candidate questions, the Solver checks logical soundness and solvability, the Educator evaluates alignment with the educational objective, and the Checker performs final verification. This process can reduce the time cost of designing diverse questions while helping teachers maintain control over pedagogical quality. For adaptive learning systems, such generation may also provide questions with stronger diagnostic potential, because varied questions under the same objective can reveal students' understanding from different perspectives.

\subsection{Applicability and Limitations}

The results also clarify the applicability and boundaries of EduAgentQG. The constructed dataset covers Grades 1--9, 634 knowledge concepts, 16 core competency categories, and three difficulty levels, and the evaluation is conducted on both MathChoice and MathBlank. This setting allows the framework to be tested under different question formats, where multiple-choice questions require plausible options and fill-in-the-blank questions require precise formulation and answer uniqueness. EduAgentQG also shows stable improvements across Gemini-2.5-Flash, GPT-4o-mini, and Qwen2.5-72B-Instruct, suggesting that the observed gains mainly come from the framework-level design rather than from a single backbone model. In addition, the human evaluation results support the reliability of the evaluation protocol. The agreement between human experts and the LLM ensemble indicates that, when evaluation dimensions are clearly defined, LLM-as-judge can provide useful support for large-scale evaluation of generated questions. Nevertheless, the empirical results of this study are limited to mathematics question generation. Mathematics questions usually have clearer answer constraints and more explicit solution paths than questions in subjects involving open-ended responses, reading comprehension, explanation, or writing. Therefore, although the proposed workflow reflects general design principles for educational question generation, extending EduAgentQG to other school subjects may require subject-specific evaluation rubrics, revised judging criteria, and additional validation. EduAgentQG also introduces additional costs because it requires multiple agents, candidate generation, evaluation, and rewriting. In practical deployment, the number of generation directions, candidate questions, and rewriting steps should be adjusted according to the application scenario.

\section{Conclusion}
In this work, we propose EduAgentQG, a multi-agent collaborative framework for personalized mathematics question generation under explicit educational objectives. EduAgentQG decomposes each educational objective into structured design plans and coordinates five specialized agents through an iterative process of planning, generation, evaluation, and refinement. By integrating diversity-oriented planning, dimension-wise evaluation, and closed-loop feedback, the framework improves both the controllability and pedagogical alignment of generated questions. Experiments on a multi-dimensional mathematics question generation benchmark demonstrate that EduAgentQG consistently outperforms the compared baselines in diversity, Objective Consistency, and Win Rate across different question formats and backbone models. Human evaluation and ablation studies further verify the practical utility of EduAgentQG and the contribution of its key components.

\section*{Acknowledgements}
This work was supported in part by "Pioneer" and "Leading Goose" R\&D Program of Zhejiang (2025C02037) and Key R\&D Program of Hangzhou (2025SZDA0254). This work was supported by the National Natural Science Foundation of China (Grant No.~62477012), the Natural Science Foundation of Shanghai (Grant No.~23ZR1418500), and the AI for Science Program of the Shanghai Municipal Commission of Economy and Informatization (Grant No.~2025-GZL-RGZN-BTBX-01014).

\section*{Data availability}

The datasets and source code associated with this study are publicly available in the GitHub repository: 
\url{https://anonymous.4open.science/r/EduagentQG-D2B0}.

\bibliographystyle{cas-model2-apa-v2}
\bibliography{ref}

\appendix
\appendix
\section{LLM-as-Judge Prompts}
\label{app:judge_prompt}

To make the automatic evaluation process more transparent and reproducible, we provide the prompts used for LLM-as-judge evaluation. The prompts are used for Objective Consistency evaluation and Win Rate evaluation, respectively. Each prompt specifies the input information, the evaluation dimensions, the scoring or preference rules, and the required output format for the LLM judges. For Objective Consistency, the prompt asks the judge to evaluate grade-level alignment, knowledge concept alignment, difficulty alignment, core competency alignment, and solvability. For Win Rate, the prompt asks the judge to compare the generated question with the gold-standard question under the same educational objective.

\begin{tcolorbox}[
colback=gray!5,
colframe=gray!70,
title={Prompt for Objective Consistency Evaluation},
breakable
]
You are a senior expert in mathematics education, curriculum design, and mathematics assessment. 
Please treat the following task as a formal mathematics education evaluation task.

You will be given a mathematics question and its corresponding educational objective. 
The educational objective includes the target grade level, knowledge concept, difficulty level, and core competency requirement. 
Your task is to evaluate how well the given question is consistent with the educational objective.

Please evaluate the question according to the following five dimensions:

\begin{itemize}
    \item \textbf{Grade-Level Alignment}: 
    Check whether the question is appropriate for the target grade level specified in the educational objective. 
    A lower score should be assigned if the question requires knowledge, language, or a solution process that is clearly above or below the target grade level.

    \item \textbf{Knowledge Concept Alignment}: 
    Check whether the question fully covers the knowledge concept required by the educational objective. 
    A lower score should be assigned if the question misses the target knowledge concept or introduces irrelevant mathematical concepts.

    \item \textbf{Difficulty Alignment}: 
    Analyze whether the difficulty level of the question matches the target difficulty specified in the educational objective. 
    Consider the cognitive demand, reasoning complexity, computation load, and solution steps. 
    A lower score should be assigned if the question is clearly too easy or too difficult.

    \item \textbf{Core Competency Alignment}: 
    Analyze whether the question supports the development or assessment of the mathematical competency required by the educational objective, such as mathematical reasoning, problem solving, modeling ability, spatial imagination, or computational ability. 
    A lower score should be assigned if the question only involves mechanical computation and does not reflect the intended competency requirement.

    \item \textbf{Solvability}: 
    Determine whether the question is clearly stated, logically coherent, and solvable with sufficient conditions. 
    If an answer is provided, verify whether the answer is correct. 
    A lower score should be assigned if the question is ambiguous, has missing conditions, is unsolvable, has multiple possible answers, or contains an incorrect answer.
\end{itemize}

For each dimension, assign a score from 0 to 10. 
The score may contain one decimal place, such as 8.2. 
Higher scores indicate stronger consistency with the educational objective.

Please remain objective and rigorous throughout the evaluation. 
Only use the given educational objective and question as the basis for judgment, and do not introduce additional assumptions.

\textbf{Input:}

Question: \\
\{QUESTION\}

Educational Objective: \\
\{EDUCATION\_GOALS\}

\textbf{Output:}

Please strictly output the result in JSON format without any additional text or code block:

\begin{verbatim}
{
  "que_id": "{QUESTION_ID}",
  "evaluation_result": {
    "grade_level_alignment": {
      "score": 0.0,
      "reason": ""
    },
    "knowledge_concept_alignment": {
      "score": 0.0,
      "reason": ""
    },
    "difficulty_alignment": {
      "score": 0.0,
      "reason": ""
    },
    "core_competency_alignment": {
      "score": 0.0,
      "reason": ""
    },
    "solvability": {
      "score": 0.0,
      "reason": ""
    }
  },
  "educational_objective": "{EDUCATION_GOALS}",
  "question": "{QUESTION}"
}
\end{verbatim}
\end{tcolorbox}

\begin{tcolorbox}[
colback=gray!5,
colframe=gray!70,
title={Prompt for Win Rate Evaluation},
breakable
]
You are a senior expert in mathematics education, curriculum design, and mathematics assessment. 
Please carefully compare the following two mathematics questions under the same educational objective.

Question 1 is the gold-standard question, and Question 2 is the model-generated question. 
Your task is to determine which question better satisfies the given educational objective. 
If Question 2 is overall superior to Question 1 in achieving the educational objective, select Question 2. 
Otherwise, select Question 1.

Please compare the two questions according to the following dimensions:

\begin{itemize}
    \item \textbf{Completeness of Concept Coverage}: 
    Analyze whether each question fully covers the knowledge concept required by the educational objective. 
    Check whether there are missing core concepts, insufficient coverage, or redundant irrelevant content.

    \item \textbf{Appropriateness of Difficulty Level}: 
    Evaluate whether each question matches the target difficulty level and is appropriate for the target grade level.

    \item \textbf{Relevance to Ability Development}: 
    Determine whether each question is consistent with the target students' ability level and the competency requirement described in the educational objective.

    \item \textbf{Contribution to Mathematical Literacy}: 
    Analyze whether each question helps assess or develop the intended mathematical literacy, such as mathematical reasoning, problem solving, modeling ability, spatial imagination, or computational ability.

    \item \textbf{Soundness of Structural Design}: 
    Evaluate whether the structure of each question is reasonable, including the organization of information, condition setting, guidance quality, and overall problem design.

    \item \textbf{Clarity and Coherence of Text}: 
    Assess whether the wording of each question is clear, concise, and coherent, and whether it effectively conveys the information needed for problem solving without ambiguity or misleading hints.

    \item \textbf{Solvability}: 
    Determine whether each question is solvable, whether the conditions are sufficient, and whether the logic is rigorous. 
    If an answer is provided, verify whether the answer is correct.
\end{itemize}

Please make an overall judgment based on the above dimensions. 
Do not compare the model-generated question with other generated questions. 
Each comparison should be made independently. 
If the two questions have different strengths and weaknesses, select the one that better achieves the educational objective overall.

\textbf{Input:}

Educational Objective: \\
\{EDUCATION\_GOALS\}

Question 1: \\
\{QUESTION\_1\}

Question 2: \\
\{QUESTION\_2\}

\textbf{Output:}

Please strictly output the result in JSON format without any additional text or code block:

\begin{lstlisting}
{
  "Better_Quest": 1,
  "reason": "Please provide a detailed explanation of why the selected question is better and indicate the dimensions in which it performs better."
}
\end{lstlisting}
\end{tcolorbox}

\begin{table*}[tb]
\centering
\footnotesize
\setlength{\tabcolsep}{5pt}
\renewcommand{\arraystretch}{1.20}

\caption{Representative examples of annotated mathematics questions in the dataset.}
\label{tab:annotation_examples}

\begin{tabularx}{\textwidth}{
@{}
C{0.035\textwidth}
Y
L{0.30\textwidth}
@{}
}
\toprule
\textbf{ID}
&
\textbf{Question}
&
\textbf{Annotations}
\\
\midrule

Q1
&
After a rope is folded in half three times, each resulting segment is what fraction of the original length?
\newline
A.~$\frac{1}{2}$ \quad
B.~$\frac{1}{3}$
\newline
C.~$\frac{1}{6}$ \quad
D.~$\frac{1}{8}$
&
\textbf{Grade:} 3
\newline
\textbf{Knowledge Concept:} Introduction to Fractions
\newline
\textbf{Difficulty:} Easy
\newline
\textbf{Competency Label:} Number sense
\newline
\textbf{Answer:} D
\\
\addlinespace[6pt]

Q2
&
An isosceles triangle has a perimeter of $21$ cm and a base of $5$ cm. The length of each equal side is \underline{\hspace{0.8cm}} cm.
&
\textbf{Grade:} 4
\newline
\textbf{Knowledge Concept:} Measuring the Perimeter of a Triangle
\newline
\textbf{Difficulty:} Medium
\newline
\textbf{Competency Labels:} Geometric intuition; Basic ability
\newline
\textbf{Answer:} $8$ cm
\\
\addlinespace[6pt]

Q3
&
A cube-shaped iron block with an edge length of $4$ cm is completely submerged in a container filled to the brim. What is the volume of the water that overflows?
\newline
A.~$16$ mL \quad
B.~$8$ mL
\newline
C.~$64$ mL \quad
D.~$96$ mL
&
\textbf{Grade:} 5
\newline
\textbf{Knowledge Concept:} Capacity and Units of Capacity
\newline
\textbf{Difficulty:} Hard
\newline
\textbf{Competency Labels:} Quantity sense; Geometric intuition; Computation
\newline
\textbf{Answer:} C
\\
\addlinespace[6pt]

Q4
&
The largest possible circle is drawn inside a square with a side length of $6$ cm. The diameter of the circle is \underline{\hspace{0.6cm}}, its radius is \underline{\hspace{0.6cm}}, its circumference is \underline{\hspace{0.6cm}}, and its area is \underline{\hspace{0.6cm}}.
&
\textbf{Grade:} 6
\newline
\textbf{Knowledge Concept:} Circles: Review and Summary
\newline
\textbf{Difficulty:} Medium
\newline
\textbf{Competency Labels:} Spatial concept; Reasoning; Model awareness
\newline
\textbf{Answer:} $6$ cm; $3$ cm; $18.84$ cm; $28.26$ cm$^{2}$
\\
\addlinespace[6pt]

Q5
&
Suppose $\angle A$ and $\angle B$ are complementary, while $\angle B$ and $\angle C$ are supplementary. If $\angle C=150^\circ$, then $\angle A$ is:
\newline
A.~$30^\circ$ \quad
B.~$45^\circ$
\newline
C.~$60^\circ$ \quad
D.~$75^\circ$
&
\textbf{Grade:} 7
\newline
\textbf{Knowledge Concept:} Introduction to Geometry: Review and Summary
\newline
\textbf{Difficulty:} Medium
\newline
\textbf{Competency Labels:} Geometric intuition; Computation; Reasoning
\newline
\textbf{Answer:} C
\\

\bottomrule
\end{tabularx}
\end{table*}

\begin{table*}[tb]
\centering
\small
\caption{Definitions of the 16 core competency categories used in the dataset annotation.}
\label{tab:competency_definitions}
\begin{tabularx}{\textwidth}{p{0.25\textwidth}X}
\toprule
\textbf{Core competency} & \textbf{Definition} \\
\midrule
Computation & The ability to correctly perform mathematical operations, calculations, transformations, and procedural solution steps. \\
Geometric intuition & The ability to understand, represent, and reason about geometric figures, shapes, positions, and visual structures. \\
Basic ability & Fundamental mathematical skills required for understanding and solving elementary mathematical problems, including basic computation, recognition, and application of simple rules. \\
Spatial concept & The ability to understand spatial relationships, directions, positions, movement, and orientation in two-dimensional or three-dimensional contexts. \\
Application awareness & The ability to connect mathematical knowledge with real-life situations and apply mathematical methods to solve practical problems. \\
Quantity sense & The ability to understand quantities, units, measurement relationships, and the relative size or magnitude of quantities. \\
Number sense & The ability to understand numbers, numerical relationships, place value, estimation, and numerical representations. \\
Reasoning awareness & The awareness of using logical relationships, evidence, and mathematical rules to justify conclusions or solution processes. \\
Symbol awareness & The ability to understand and use mathematical symbols, expressions, formulas, and symbolic representations. \\
Creativity & The awareness of exploring alternative strategies, flexible thinking, and non-routine approaches in mathematical problem solving. \\
Data awareness & The ability to read, organize, compare, and interpret data presented in tables, charts, or statistical contexts. \\
Model awareness & The awareness of representing real-world or mathematical situations using simplified mathematical structures, relationships, or models. \\
Model concept & The ability to understand, construct, and interpret mathematical models that describe quantitative or structural relationships. \\
Reasoning & The ability to conduct logical inference, analyze relationships, justify conclusions, and complete multi-step mathematical reasoning. \\
Abstraction ability & The ability to extract mathematical structures, patterns, or relationships from concrete situations and express them in general or symbolic forms. \\
Data concept & The ability to understand basic statistical concepts, data representations, data distributions, and relationships among data elements. \\
\bottomrule
\end{tabularx}
\end{table*}

\section{Examples of Dataset Annotations}
\label{app:annotation_examples}

The dataset comprises school-level mathematics questions annotated according to a structured educational objective. Each instance contains the question text and its corresponding answer, together with labels for grade level, knowledge concept specification, difficulty level, and core competency specification. The final annotation of each question is represented as $(G, K, D, S)$, where $G$ denotes the grade level, $K$ denotes the knowledge concept specification, $D$ denotes the difficulty level, and $S$ denotes the core competency specification. The knowledge concept specification may contain one or more knowledge concepts, whereas the core competency specification may involve either a single competency category or a combination of multiple competency categories.

More specifically, the grade-level label indicates the curriculum stage for which the question is intended, while the knowledge concept specification identifies the mathematical content required to solve it. The difficulty label reflects the expected cognitive and reasoning demand, such as the number of solution steps, the degree of condition transformation, and the extent of concept integration. The core competency specification identifies the mathematical abilities reflected by the question, such as Computation, geometric intuition, Reasoning, number sense, or application awareness. Multiple core competency categories are assigned only when each category represents an essential component of the intended solution process or assessment objective.

These annotations characterize not only the mathematical content required to solve a question, but also its expected cognitive and reasoning demand and the mathematical abilities required by the question. In this way, each question is linked to an explicit and fine-grained educational objective rather than being described only by a topic-level label. This annotation protocol provides the structured educational information used in the question-generation and objective-aware evaluation processes.

Table~\ref{tab:annotation_examples} presents five representative examples to illustrate the annotation protocol. The examples cover different grade levels, question formats, knowledge concepts, difficulty levels, and core competency categories, including Introduction to Fractions, Measuring the Perimeter of a Triangle, Capacity and Units of Capacity, Circles: Review and Summary, and Introduction to Geometry: Review and Summary. They illustrate how the four components of the structured educational objective are jointly instantiated in individual dataset records. In particular, the examples show that a question may be associated with either a single core competency category or multiple core competency categories when more than one competency dimension is essential to solving the question. The examples are intended to provide a concrete illustration of the annotation protocol rather than an exhaustive summary of the dataset distribution.

\section{Definitions of Core Competency Categories}

This appendix provides the definitions of the 16 core competency categories used in the dataset annotation. These categories specify the core competency dimension reflected by a question and are used together with grade level, knowledge concept, and difficulty level to form the structured educational objective of each instance. The competency categories describe the mathematical abilities that a question is intended to assess, such as number sense, Computation, geometric intuition, Reasoning, and application awareness. They therefore complement the knowledge concept specification by indicating not only what mathematical content is involved, but also what type of mathematical ability the question is designed to develop or evaluate.

During annotation, each question was assigned either a single core competency category or multiple core competency categories, depending on the competencies that were essential to solving the question. The core competency categories were not treated as mutually exclusive. For example, a geometry question may require geometric intuition to interpret spatial relationships and Reasoning to derive a conclusion. Similarly, a real-world mathematics question may involve application awareness because it requires translating a practical situation into a mathematical representation, as well as Computation because it requires appropriate calculations or procedures.

To ensure that the core competency specification remained focused and interpretable, multiple competency categories were retained only when each category represented an essential component of the question's intended solution process or assessment objective. A competency category was not assigned merely because it was incidentally involved in solving the question. In other words, a question received multiple competency categories only when removing one of them would substantially change the mathematical ability reflected by the question.

When annotators assigned different core competency categories to the same question, the discrepancy was reviewed by a third annotator. The final annotation was determined according to the definitions provided below and the dominant mathematical ability required by the question. In particular, the adjudication considered the primary cognitive and reasoning demand of the question, the main solution process, and the core competency dimension most clearly reflected by the question. Therefore, multi-label annotations indicate that multiple core competency categories are essential to the question, whereas conflicting initial annotations were resolved through adjudication and were not retained as separate final labels.

\end{document}